\documentclass[11pt,a4paper]{article}
\usepackage{jheppub}
\def\S2{\bar{S}}
\def\a{\alpha}

\def\and{a_{n}^\dagger}

\def\sn2d{\Sn2^\dagger}

\def\({\left(}
\def\){\right)}
\def\<{\left\langle}
\def\>{\right\rangle}

\newcommand\ee{\end{eqnarray}}      
\newcommand\be{\begin{eqnarray}}
\newcommand\ba{\begin{array}}           
\newcommand\ea{\end{array}}
\newcommand\eeq{\end{equation}}     
\newcommand\beq{\begin{equation}}

\title{Left/right entanglement and thermalization of time dependent plane wave Green-Schwarz superstring}

\author{D\'afni F. Z. Marchioro,}

\author{Daniel Luiz Nedel}

\affiliation{Universidade Federal da Integra\c{c}\~ao Latino-Americana \\
Avenida Tancredo Neves 6731, Foz do Igua\c{c}u, Brasil}

\emailAdd{dafni.marchioro@unila.edu.br}
\emailAdd{daniel.nedel@unila.edu.br}

\abstract{In this work we study new issues  involving the type IIB superstring in a time dependent plane wave background with a constant self-dual Ramond-Ramond 5-form  and a linear dilaton in the light-like direction.  We construct a unitary Bogoliubov generator which relates the asymptotically flat superstring Hilbert space to the finite time Hilbert space. The time dependent vacuum is
a superposition of $SU(1, 1)\times SU(2)$ coherent states, which has a particular structure of excitation, characterized by a condensation of
right and left moving supertring modes.  We calculate the time dependent left/right entanglement entropy and carry out the summation over the oscillator modes of the superstring two-point function. We show that, close to the null singularity, the entanglement entropy is well-behaved. In particular, for asymptotically flat observers, the closed superstring vacuum close to the singularity appears as superstring thermal vacuum, which is unitarily inequivalent to the asymptotically flat vacuum. Actually, we show that close to the singularity the superstring thermalizes and the entanglement entropy becomes a thermodynamical entropy for a supersymmetric two-dimensional gas.}


\arxivnumber{0000.0000}

\dedicated{Dedicated to Yasmim Marchioro Nedel.}

\begin{document}

\maketitle

\flushbottom

\section{Introduction}

One of the important open problems in high energy physics is the correct treatment of time dependent backgrounds. It is generally hoped that string theory has the tools to solve the conceptual and technical problems associated to quantization in time dependent geometries; in particular, string theory may help to understand the nature of space-like singularities.   Despite all the progress that has been made in the last years involving orbifolds \cite{orb1, orb2, orb3, orb4, orb5, orb6, orb7, orb8, orb9}, matrix theory \cite{matr1, matr2, matr3, matr4}, tachyon condensation \cite{tac1, tac2, tac3, tac4}, among other ideas\footnote{For a review, see, for example, \cite{berkoos}.}, the quantum properties of superstring sigma model in general  time dependent background, in particular with non zero Ramond-Ramond fluxes, still remain elusive, mainly owing to the breakdown of the string perturbative techniques close to singularity.
 
 In order to understand superstring theory at time dependent background and cosmological singularities, we need to answer some questions. Maybe the first question in quantum cosmology is whether time simply begins and ends - the string pre-Big Bang scenario potentially answers this question \cite{pbb}. In this scenario quantum effects may produce a kind of bounce with a semi-classical spacetime on the other side. Another question is the nature of the null closed curves which appear in time dependent orbifolds, and how to take into account the string-winding effects associated with these curves. Finally (but not the last), in a general time-dependent background there is no natural definition of the vacuum. As a consequence, it is not always clear what are the correct observables of string theory.

A complete program to answer the questions above needs to take into account the $\alpha'$ correction, the $g_s$ correction and the non-perturbative sector of string theory. On the other hand, as usual in theoretical physics, we can get insights by studying models that are simple enough to have, for example, non-perturbative effects over control, but complicated enough to illustrate nontrivial effects. Although the employing of perturbative methods to treat superstrings in time independent backgrounds in the case of time dependent ones is not straightforward, there are some particular time dependent backgrounds which it is. This is the case of time dependent plane wave background.

The study of string theory in a plane wave background has a long and interesting  history since the pioneering works of references \cite{HS1, HS2, RB, VS1, VS2, VS3}.  Concerning the time independent case, the plane wave background with maximal supersymmetry \cite{BFHP1, BFHP2, BFP} has been produced by the Penrose limit \cite{RP,RG} on the $AdS_5 \times S^5$ solution of type IIB superstring theory. The string sigma model in this time independent background is exactly solvable \cite{RM,MT}, so that it provides an example of AdS/CFT correspondence  beyond the supergravity approximation - the so-called BMN correspondence \cite{BMN}. Later on, a time dependent plane wave bosonic string model was studied in \cite{PRT}. For a particular choice of the metric's parameter, this time dependent geometry can be obtained through the Penrose limit of a cosmological, Dp-brane or fundamental string background. It has a null cosmology interpretation - in particular, the model admits a pre-Big Bang phase scenario. In the light-cone gauge, the string's equation of motion has been solved and it is argued that the string passes through the null singular point, although there is a discontinuity in the time derivative in the zero mode sector. In this case, the string coupling close to singularity remains small.  In \cite{Bin}, the type IIB Green-Schwarz superstring in a plane wave time dependent background with constant Ramond-Ramond flux was quantized. This model also has a null cosmology interpretation but does not allow a pre-Big Bang phase scenario. It was shown that the spectrum of the bosonic and fermionic excitations is symmetric  and the zero-point energy cancels between the bosonic and fermionic sectors; however, the string coupling is very strong near the Big Bang singularity. Moreover, as in the model studied in \cite{PRT}, there is an asymptotically flat limit. In the present work,  we are going to study new issues involving this time dependent superstring sigma model, in particular concerning left/right entanglement, two-point functions and thermalization.

The main characteristic of time dependent plane wave superstring models lies in the fact that, although there is no particle creation in the background, there is string mode creation. The core of this effect is the presence of a time dependent mass in the worldsheet model. This implies that the worldsheet vacuum is not unique and we can construct a Bogoliubov operator to map different representations of the Poisson algebra. In particular, we have related the asymptotically flat vacuum with a time dependent vacuum. As a consequence of the Bogoliubov transformation,  the time dependent vacuum is a left/right superstring entanglement state. Interesting to note that this entanglement is produced by the background. In general, the left/right entanglement state of conformal theories is a linear combination of Ishibashi states and it is not normalized \cite{DipDas}. Here, due to the unitarity of the Bogoliubov transformation, the time dependent superstring entanglement state is normalized. 

An important object in field and string theory is the two-point function. Although the use of string's perturbative techniques is only possible for this model close to flat space limit, it is important to have knowledge about the dependencies of the background field and the worldsheet spacetime structure of string propagators. To this end, it is necessary to carry out the mode summation for the string
two-point function and present it in terms of analytic functions. This was done at \cite{Ryang} for the bosonic model discussed at \cite{PRT}. The same analysis is done here for the type IIB Green-Schwarz superstring in a plane wave time dependent background with constant Ramond-Ramond flux. In particular, the short-distance behavior is analyzed close to the flat space and close to the singularity, for both vacuum states. In the work of reference \cite{Ryang}, an approximation is used to go to the continuous limit and to write the sums involving the two-point function as integrals. For the model studied here, this approach is not possible and we need to deal directly with the sums. A perturbative approach is used to carry out the two-point mode summation  and write it in terms of Hypergeometric and q-Polygamma functions for the time independent vacuum, and in terms of modified Bessel functions for the time dependent left/right entanglement vacuum.

Once we have a left/right entanglement state, a natural step to take is to investigate the effects of tracing over the left-moving degrees of freedom and to calculate the entanglement entropy. Usually, the entanglement entropy is defined as the von Neumann entropy corresponding to the reduced density matrix  $S_E=-Tr \rho_A\ln\rho_A$, where the reduced density matrix $\rho_A$  of a subspace of the Hilbert space $\mathcal H$ is obtained by tracing out the degrees of freedom living in its complement $\mathcal H_B$. It has to be emphasized that the Hilbert space is not geometrically partitioned here - the division of the system into subsystems $A$ and $B$ does not follow the more traditional geometric delimitation. Instead, as a consequence of the Bogoliubov transformation, the Hilbert space is decomposed into string's left- and right-moving degrees of freedom. Therefore, this entanglement is more related to the concept of momentum entanglement, investigated in \cite{Balasu}.\footnote{This kind of entanglement was studied for the first time in string theory in \cite{GMN} and later on in \cite{z1, z2}.} On the other hand, once we show that the entanglement state can be also generated by an entropy operator, the entanglement entropy has the form of a thermodynamical entropy.

With the entanglement state being produced by the background and the entropy, the natural question that arises is what happens at the cosmological singularity. As for this model the string coupling gets bigger as the string is closer to the singularity, any question involving how the string resolves or passes through the spacelike singularity needs non perturbative information to be answered. Surprisingly, although the Hamiltonian of the superstring diverges at the singularity, the entanglement state and entropy are well behaved. In fact, we show that, as the string approaches the singularity as seen by asymptotic observers, the left/right entanglement entropy becomes the thermodynamical entropy for a 2d supersymmetric gas and the entanglement state becomes a thermal state. As the left-moving degrees of freedom have been traced out, the entropy has the form of an open string entropy, as well as the thermal state. Again, this is a consequence of the time dependent mass.\footnote{It is well-known that, even in a free field theory, when we suddenly change the mass, the resulting real time correlation functions become ``thermal'' at late time \cite{Calabre}, \cite{Rigol}.} Important to note that this worldsheet thermalization does not occur in the model studied in \cite{PRT} - this is a particularity of this Ramond-Ramond background.

This article is divided as follows: in Section 2 the model is presented, following the reference \cite{Bin}; in Section 3, we construct a unitary Bogoliubov generator which relates the asymptotically flat string Hilbert space to the finite time Hilbert space; in Section 4 the two-point function is calculated and we show a different behavior for the two vacuum states; in Section 5 we explore the fact that the time dependent vacuum is a time dependent left/right entanglement state and calculate the time dependent  entanglement entropy; in Section 6 we show the thermalization of the system as the string approaches the singularity; the conclusions are presented in Section 7; lastly, in the Appendix we present the so-called Liouville-von Neumann (LvN) approach \cite{Kim,KMMS,Lewis}
 to study the non equilibrium quantum dynamics of time dependent systems, which has been used to show the zero mode thermalization.

\section{The model}

Consider the type IIB Green-Schwarz (GS) superstring in the following time dependent background with Ramond-Ramond flux

\begin{eqnarray}
ds^2 = -2dx^+ dx^- -\lambda(x^+)\,x_I^2\,dx^+ dx^+ +dx^Idx^I\,,\nonumber \\
\phi=\phi(x^+)\,,\quad\quad(F_5)_{+1234}=(F_5)_{+5678}=2f.
\label{BG}
\end{eqnarray}

\noindent where $\phi$ is the dilaton and $F_5$ the Ramond-Ramond field. Here the supersymmetry preserved by the background is reduced from maximal (32 supercharges) to 1/2 (16 supercharges), as usual for a generic plane wave.

The worldsheet action is 

\begin{equation}
S=S_B+S_F \:,
\end{equation}

\noindent where 

\begin{eqnarray}
S_B&=&-\frac{1}{4\pi \alpha'}\int d^2\sigma\sqrt{-g}\,g^{ab} G_{\mu\nu} \partial_{a}x^{\mu}\partial_{b}x^{\nu}\,\nonumber\\
&=&-\frac{1}{4\pi \alpha'}\int d^2 \sigma\sqrt{-g}\,g^{ab}(-2\,\partial_a x^{+}\partial_b x^{-} -\lambda\, x^2_{I}\partial_a x^{+}\partial_b x^{+}+\partial_a x^{I}\partial_b x^{I})\,,
\label{BosA}
\end{eqnarray}

\noindent and

\begin{eqnarray}
S_F=-\frac{i}{2\pi\alpha'}\int d^2\sigma
(\sqrt{-g}g^{ab}\delta_{AB}-\epsilon^{ab}\sigma_{3AB})\,\partial_{a}x^{\mu}\,
\bar{\theta}^A\Gamma_{\mu}(\hat{D}_b\theta)^B + {\mathcal O} (\theta^3) \,,
\label{FerA}
\end{eqnarray}

\noindent are the bosonic and fermionic string actions, respectively, and

\begin{eqnarray}
\sigma_3 = \mbox{diag}(1,-1)\:, \nonumber\\
\hat{D}_b=\partial_b+\Omega_\nu\,\partial_{b}x^\nu \:,
\end{eqnarray}

\noindent with $\hat{D}_b$ being the pull-back of the covariant derivative to the worldsheet. The indices $a,b$ are worldsheet indices, $A,B = 1,2$ and $\mu$ is the spacetime index. The spin connection $\Omega_\nu$ is defined by

 \begin{eqnarray}
\Omega_-&=&0,\nonumber\\
\Omega\,_I&=&\frac{i\,e^\phi}{4}f\,\Gamma^+(\Pi+\Pi')\,\Gamma_I\,\sigma_2,\nonumber\\
\Omega_+&=&-\frac{1}{2}\lambda\,x^I\Gamma^{+I}\textbf{1}+
\frac{i\,e^\phi}{4}f\,\Gamma^+(\Pi+\Pi')\,\Gamma_+\sigma_2 \:,
\label{CovDiff}
\end{eqnarray}

\noindent with $\Pi = \Gamma^{1}\Gamma^{2}\Gamma^{3}\Gamma^{4} = \mbox{diag} ({\mathbf 1}_4, -{\mathbf 1}_4)$, $\Pi' =  \Gamma^{5}\Gamma^{6}\Gamma^{7}\Gamma^{8}$, $\Gamma^{\pm} = (\Gamma^0 \pm \Gamma^9)/\sqrt{2}$ and $\sigma_2$ is the Pauli matrix. We are omitting higher orders in $\theta$, which do not contribute after using the light-cone gauge \cite{Metsaev},\cite{Sadri}. The representation of $\Gamma$-matrices chosen is such that $\Gamma^0=C$ where $C$ is the 10d charge conjugation; therefore, the components of $\theta^{A}$ are all real. The $\theta^A$ are 10d spinors, that is, $\theta^A_\alpha$  with $\alpha=1,2,\ldots,16\,$, and $A=1,2$.  

Conformal invariance of the worldsheet demands
 
\begin{eqnarray}
R_{\mu\nu}=-2D_{\mu}D_{\nu}\phi+\frac{1}{24}e^{2\phi}(F^2_{5})_{\mu\nu}.
\label{Conformal}
\end{eqnarray}

\noindent Because the only non zero component of the Ricci curvature tensor $R_{\mu\nu}$ with respect to the metric is $R_{++}=8\lambda(x^+)$, when we put (\ref{BG}) into (\ref{Conformal}) we find

\begin{eqnarray}
\lambda=-\frac{1}{4}\phi''+f^2e^{2\phi}\,.
\end{eqnarray}

\noindent Notice that $e^{\phi} = g$ is the string coupling.  

Let us analyze some cases that have been studied in \cite{Bin} and \cite{PRT}. If we turn off the Ramond-Ramond flux,  we can choose

\begin{equation}
\phi =\phi_0 -\frac{1}{2}d\lambda_0 (x^+)^2,\:\: \lambda_0= \mbox{constant} > 0 \:.
\label{btsey}
\end{equation}

\noindent In this case the string coupling is $g= g_0e^{-\frac{1}{2}d\lambda_0 (x^+)^2} $ and $\lambda= \frac{k}{(x^+)^2}$.  The metric admits a null cosmology interpretation with a cosmological singularity at $x^+ =0$ and a pre-Big Bang phase ($x^+ <0$). Note that, even at the singularity, the string coupling remains small if $g_0$ is small, and in the asymptotic limits ($x^+ =\pm \infty$) the metric is flat. However, after fixing the kappa symmetry in  the GS superstring ($\Gamma^+ \theta^A=0$), only the second term in $\Omega_+$ has non vanishing contributions in the fermionic part. This goes to zero if the Ramond-Ramond field is zero. 

A solution of (\ref{Conformal}) with non zero constant Ramond-Ramond field ($f=f_0$) is:

\begin{equation}
\phi= -c x^+,\:\: \lambda= f_0^2e^{-2cx^+} \:.
\label{modelo}
\end{equation}

\noindent In this case, the metric also admits a null cosmology interpretation but the cosmological singularity is located in $x^+= -\infty$. In this model the string coupling $g$ diverges at the singularity. This is the model that we are going to explore in this work. 

\subsection{The light-cone superstring action}

 For the model (\ref{modelo}), we fix the gauge symmetries choosing light-cone gauge
 
\begin{eqnarray}
 x^+ = \alpha' p^+ \tau \:, \:\:\: p^+ > 0 \:, \nonumber\\ 
\Gamma^+ \theta^A = 0 \:,
\label{gfix}
\end{eqnarray}

\noindent and

\begin{equation}
\sqrt{-g} g^{ab} = \left(
\begin{array}{lr}
-1 & 0 \\
0 & 1
\end{array}
 \right) \:.
 \end{equation}
 
 \noindent The second equation in (\ref{gfix}) implies
 
\begin{eqnarray}
(\theta^A)^T \Gamma^I \theta^B = 0, \:\: \forall A, B \:, \nonumber\\
(\Omega_I)^{A}_{\:\:B} \theta^B = 0 \:, \nonumber\\
\Pi \theta^A = \Pi' \theta^A \:.
\label{kappa}
\end{eqnarray}

 As usual, after fixing the kappa symmetry ($\Gamma^+ \theta^A = 0$), the ten dimensional fermions are reduced to $SO(8)$ representation. Since  the ten dimensional $\theta^1_{\alpha}$ and $\theta^2_{\alpha}$ spinors  have the same chirality, both of them end up to be in the same $SO(8)$ fermionic representation. Moreover, after the gauge fixing, the only term of the spin connection that contributes is $\Omega_+$, and from the $\partial_a X^{\mu}\Gamma_{\mu}$ term in the fermionic action, only $\partial_a X^{+}\Gamma_{+}$ contributes. Finally, the light-cone GS superstring action is
 
\begin{equation}
S= S_B+S_F,
\end{equation}

\noindent where 

\begin{eqnarray}
S_B&=&\frac{1}{4\pi \alpha'}\int d\tau \int_0^{2\pi}\hspace{-0.2cm}d\sigma
\,\,(\partial_{\tau}x^I\partial_{\tau}x^I-
\partial_{\sigma}x^I\partial_{\sigma}x^I-\tilde{f}^2e^{-2\,\tau}x^2_I) \:,
\label{BA}
\end{eqnarray}  
 
\noindent is the bosonic part and

\begin{equation}
S_F=\frac{i p^+}{\sqrt{2}\pi}\int
d\tau\int_0^{2\pi}\hspace{-0.2cm}d\sigma\,
(\theta^{1T}\partial_\tau\theta^1+\theta^{2T}\partial_\tau\theta^2
+\theta^{1T}\partial_\sigma\theta^1-\theta^{2T}\partial_\sigma\theta^2
+2\tilde{f}e^{-\tau}\theta^{1T}\Pi\theta^2) \:,
\label{FA light-cone}
\end{equation}

\noindent is the fermionic part of the action; we are omitting the fermionic indices. We have set $c=\frac{1}{\alpha'p^+}$ and  defined $\tilde{f}\equiv\alpha'p^+f_0$. Remember that now $\theta^1$ and $\theta^2$ have 8 independent components each.  In the next paragraphs the results of reference \cite{Bin} will be presented briefly.

The equation of motion for the bosonic part is

\begin{eqnarray}
(\partial^2_{\tau}-\partial^2_{\sigma}+\tilde{f}^2e^{-2\,\tau})\,x^I=0\,,
\end{eqnarray}
 
\noindent and its solution is

\begin{eqnarray}
x^I(\tau,\,\sigma)=x^I_0(\tau)+i
\sqrt{\frac{\alpha'}{2}}\sum^{\infty}_{n=1}\frac{1}{\sqrt{n}}
\left[\,T^I_n(\tau)\,e^{in\sigma}-T^I_{-n}(\tau)\,e^{-in\sigma}\right],
\label{Tn}
\end{eqnarray}

\noindent with

\begin{eqnarray}
x^I_0(\tau)&=&J_0(\tilde{f}e^{-\tau})\tilde{x}^I-\frac{\pi}{2}\alpha'\,Y_0(\tilde{f}e^{-\tau})\tilde{p}^I \:, \nonumber\\
T^I_n(\tau)&=&Z_n(\tau)\,\alpha^I_n-Z_{-n}(\tau)\,\tilde{\alpha}^I_{-n}\,,\nonumber\\
Z_n(\tau)&=&\left(\frac{\tilde{f}}{2}\right)\Gamma(1+in)J_{in}(\tilde{f}e^{-\tau})\:,
\label{Z}
\end{eqnarray}	

\noindent and $J_{m}$ and $Y_m$ are the Bessel functions of first and second kind, respectively. The reality condition on $x^I$ implies
  
\begin{eqnarray}
(\alpha^I_n)^\dag=\alpha^I_{-n}\,\,,\,\,(\tilde{\alpha}^I_n)^\dag=\tilde{\alpha}^I_{-n}\,.
\end{eqnarray}

For the fermionic action, the equations of motion are

\begin{eqnarray}
(\partial_\tau + \partial_\sigma)\theta^1
+\tilde{f}e^{-\tau}\Pi\theta^2=0,\nonumber\\
(\partial_\tau - \partial_\sigma)\theta^2
-\tilde{f}e^{-\tau}\Pi\theta^1=0.
\label{1-ord}
\end{eqnarray}

\noindent One can transform the two coupled first order equations in two decoupled second order equations as

\begin{eqnarray}
(\partial_\tau^2-\partial_\sigma^2)\,\theta^1+(\partial_\tau+\partial_\sigma)\theta^1
+\tilde{f}^2e^{-2\tau}\theta^1=0\,\,,\nonumber\\
(\partial_\tau^2-\partial_\sigma^2)\,\theta^2+(\partial_\tau-\partial_\sigma)\theta^2
+\tilde{f}^2e^{-2\tau}\theta^2=0\,\,.
\label{2-ord}
\end{eqnarray}

\noindent Note that, owing to the time dependence of the dilaton, we have a typical damping term (first time derivative of the fields); again, the equations can be resolved in terms of Bessel functions. Defining $\xi=\frac{p^+}{\sqrt{2}\,\pi}$,  the worldsheet field expansions are

\begin{eqnarray}
\theta^1(\tau,\sigma)&=&\theta^1_0(\tau)
+\sum_{n=1}^\infty \left[\,\theta_n^1(\tau)\,e^{in\sigma}+\theta_{-n}^1(\tau)\,e^{-in\sigma}\right],\nonumber\\
\theta^2(\tau,\sigma)&=&\theta^2_0(\tau)
+\sum_{n=1}^\infty \left[\,\theta_n^2(\tau)\,e^{in\sigma}+\theta_{-n}^2(\tau)\,e^{-in\sigma}\right],
\label{mode}
\end{eqnarray}

\noindent with

\begin{eqnarray}
\theta^1_0(\tau)=\frac{1}{\sqrt{4\pi\xi}}\,(\,S_0\cos({\tilde{f}e^{-\tau}})+\Pi\tilde{S}_0\sin({\tilde{f}e^{-\tau}})),\nonumber\\
\theta^2_0(\tau)=\frac{1}{\sqrt{4\pi\xi}}\,(\,\tilde{S}_0\cos({\tilde{f}e^{-\tau}})-\Pi S_0\sin({\tilde{f}e^{-\tau}})),
\label{z.m.2}
\end{eqnarray}

\begin{eqnarray}
\theta_n^1(\tau)=\frac{1}{\sqrt{4\pi\xi}}
\left(S_{n}W_n(\tau)+\Pi\tilde{S}_{-n}\,\tilde{W}^\ast_{n}(\tau)\right)\,\,,\label{T^1}\\
\theta_n^2(\tau)=\frac{1}{\sqrt{4\pi\xi}}
\left(\tilde{S}_{-n}\,W^\ast_{n}(\tau)-\Pi S_{n}\tilde{W}_n(\tau)\right)\,\,,\label{T^2}
\end{eqnarray}

\noindent and

\begin{eqnarray}
W_n(\tau)&=&\left(\frac{\tilde{f}}{2}\right)^{\hspace{-0.2cm}-in}
\hspace{-0.1cm}\Gamma\left(\frac{1}{2}+in\right)\sqrt{\frac{u}{2}}\,J_{-\frac{1}{2}+in}({\tilde{f}e^{-\tau}})\,,\\
\tilde{W}_n(\tau)&=&\left(\frac{\tilde{f}}{2}\right)^{\hspace{-0.2cm}-in}
\hspace{-0.1cm}\Gamma\left(\frac{1}{2}+in\right)\sqrt{\frac{u}{2}}\,J_{\frac{1}{2}+in}({\tilde{f}e^{-\tau}})\,.
\end{eqnarray}

\noindent The requirement that $\theta^{1,2}$ are real implies

\begin{eqnarray}
S_{-n}=S^{\dag}_{n}\,\,,\quad
\tilde{S}_{-n}=\tilde{S}^{\dag}_{n}\,\,,\quad n=0,\pm1,\pm2,\ldots \:.
\end{eqnarray}

Now let's quantize the superstring. The canonical momentum conjugate to the bosonic and fermionic coordinates are

\begin{eqnarray}
\Pi^I&=&\frac{1}{2\pi\alpha'}\,\partial_{\tau}x^I \:, \nonumber \\
\mathcal{P}^A_{\hat a} &=&\frac{i\,p^+}{\sqrt{2}\,\pi}\,\theta^A_{\hat a}\,,
\quad A=1,2\quad\mathrm{and}\quad {\hat a} =1,2,\ldots,8 \:.
\end{eqnarray}

\noindent Using the following properties of the Gamma and Bessel functions,

\begin{eqnarray}
&&\Gamma(1+in)\,\Gamma(1-in)=\frac{n\pi}{\,\sinh{n\pi}}\,\,,\nonumber\\
&&J_{\nu}(z)J'_{-\nu}(z)-J_{-\nu}(z)J'_{\nu}(z)=-\frac{2\sin{\nu\pi}}{\pi z}\,\,,
\label{Bes}
\end{eqnarray}

\noindent the bosonic canonical commutation relations

\begin{eqnarray}
[\,x^I(\tau,\sigma),\,\Pi^J(\tau,\sigma')\,]=i\,\delta^{IJ}\delta(\sigma-\sigma')\,,\label{CR}
\end{eqnarray}

\noindent imply

\begin{eqnarray}
[\tilde{x}^I,\,\tilde{p}^J]=i\,\delta^{IJ}\,\,,\,\,\,\,
[\alpha^I_n,\,\alpha^{J\dag}_m]=\delta^{IJ}\delta_{nm}\,\,,\,\,\,\,
[\tilde{\alpha}^I_n,\,\tilde{\alpha}^{J\dag}_m]=\delta^{IJ}\delta_{nm}\,\,.\label{OC}
\end{eqnarray}

\noindent For the quantization of the fermionic part, we impose the standard anticommutation relations

\begin{eqnarray}
\{\theta^A_{n{\hat a}}(\tau,\sigma)\,,\theta^B_{m{\hat b}}(\tau,\sigma')\}
=\frac{1}{2\,\xi}\,\delta^{AB}\,\delta_{n+m,0}\,\delta_{{\hat a}{\hat b}}\,\delta(\sigma-\sigma'),
\label{FAC}
\end{eqnarray}

\noindent with the other anticommutators vanishing. Using the formulas

\begin{eqnarray}
&&\Gamma \left(\frac{1}{2}+in \right)\,\Gamma\left(\frac{1}{2}-in \right)=\frac{\pi}{\,\cosh{n\pi}}\,\,,\label{GaHalf}\\
&&J_{-\frac{1}{2}+in}(z)J_{-\frac{1}{2}-in}(z)+J_{\frac{1}{2}+in}(z)J_{\frac{1}{2}-in}(z)
=\frac{2\cosh{n\pi}}{\pi z},\label{J-Hal}
\end{eqnarray}

\noindent we get

\begin{eqnarray}
\{S_{n{\hat a}},S_{m{\hat b}}\}
=\{\tilde{S}_{n{\hat a}},\tilde{S}_{m{\hat b}}\}
=\delta_{{\hat a}{\hat b}}\,\delta_{n+m,0}\,,\quad n,m=0,\pm1,\pm2,\ldots \:.
\end{eqnarray}

Although the quantization of the sigma model is straightforward, one can see that the light-cone Hamiltonian written in terms of the modes $\alpha_n$ and $S_n$ is non diagonal:

\begin{equation}
H=H_B+ H_F,
\end{equation}

\noindent where

\begin{eqnarray}
H_B&=&H_{B0}(\tau)+\frac{1}{2\alpha'p^+}
\sum^{\infty}_{n=1}\left[\,\Omega^B_n(\tau)\,
(\,\alpha^{I\dag}_{n}\alpha^I_n+\tilde{\alpha}^{I\dag}_{n}\tilde{\alpha}^I_n+1)\right.\nonumber\\
&&\hspace{4cm}\left.-C^B_n(\tau)\,\alpha^I_{n}\tilde{\alpha}^I_n
-C^{B\ast}_n(\tau)\,\tilde{\alpha}^{I\dag}_{n}\alpha^{I\dag}_{n}\,\right]\,,
\label{AC}
\end{eqnarray}

\begin{eqnarray}
H_F&=&H_{F0}(\tau)+\frac{1}{2\alpha'p^+}
\sum^{\infty}_{n=1}\left[\,\Omega^{F}_n(\tau)\,
(\,S^\dag_{n}S_n+\tilde{S}^\dag_{n}\tilde{S}_n-1)\right.\nonumber\\
&&\hspace{4cm}\left.-C^{F}_{n}(\tau)\,S_n\Pi\tilde{S}_n
-C^{F\ast}_{n}(\tau)\,\tilde{S}^\dag_{n}\Pi S^\dag_{n}\,\right]\,,
\end{eqnarray}

\begin{eqnarray}
H_{B0}(\tau)&=&
\frac{1}{2p^+}\left[(p^I_0)^2+\tilde{f}^2e^{-2\tau}\left(\frac{x^I_0}{\alpha'}\right)^2\right],\nonumber\\
H_{F0} (\tau) &=&  -\frac{2\,i}{\alpha'p^+}\,\tilde{f}\,e^{-\tau}S_0\Pi\tilde{S}_0,\label{H0}
\end{eqnarray}

\noindent and

\begin{eqnarray}
\Omega^B_n(\tau)&=&
\frac{1}{n}|\partial_{\tau}Z_n|^{2}+n\left(1+\frac{\tilde{f}^2e^{-2\tau}}{n^{2}}\right)|Z_n|^{2}\,,\nonumber\\
C^B_n(\tau)&=&
 \frac{1}{n}(\partial_{\tau}Z_n)^{2}+n\left(1+\frac{\tilde{f}^2e^{-2\tau}}{n^{2}}\right)(Z_n)^{2}\,, \nonumber \\
\Omega^{F}_n(\tau)&=&-2\,i\left[W_n\partial_{\tau}W^{\ast}_n+
\tilde{W}_n\partial_{\tau}\tilde{W}^{\ast}_n\right] \:,\nonumber \\
C^F_n(\tau)&=&-2\,i\left[W_n\partial_{\tau}\tilde{W}_n-\tilde{W}_n\partial_{\tau}W_n\,\right]. 
\label{hf}
\end{eqnarray}

\noindent Let us study the asymptotic behaviour of this Hamiltonian. Considering the asymptotic expansion of Bessel
functions as $z\rightarrow 0$,

\begin{eqnarray}
J_{\nu}(z)&\sim &\frac{1}{\Gamma(1+\nu)}\left(\frac{z}{2}\right)^{\nu}+\mathcal{O}(z^{\nu+2})\,\,,\nonumber\\
Y_0(z)&\sim&\frac{2}{\pi}\ln\frac{z}{2}\,,\,\,\,
\Omega^{F}_n(\tau)\sim2n \,,\,\,\, C^F_n(\tau)\sim0 \,,
\label{Bflat}
\end{eqnarray}

\noindent these relations imply

\begin{eqnarray}
\Omega^{F}_n(\tau)\sim 2n \,,\,\,\, C^F_n(\tau)\sim 0 \,, \nonumber\\
\Omega^{B}_n(\tau)\sim 2n \,,\,\,\, C^B_n(\tau)\sim 0 \,, 
\label{Oflat}
\end{eqnarray}

\noindent that is, the flat space result is recovered as $\tau\rightarrow+\infty$ and the Hamiltonian is diagonal. Close to the singularity, we use the following expression for $z\rightarrow \infty$

\be
J_{\nu}(z)\sim
\sqrt{\frac{2}{\pi z}}\,\cos\left(z-\frac{\nu\pi}{2}-\frac{\pi}{4}\right)\,, 
\,\,\,\,\,\,|\arg z|<\pi \,,
\label{ase}
\ee

\noindent to get

\be
\Omega^B_n(\tau)&\sim&\frac{2\cosh {n\pi}}{\sinh {n\pi}}\,\tilde{f}e^{-\tau}\,,\nonumber\\
C^B_n(\tau)&\sim&2\left(\frac{\tilde{f}}{2}\right)^{-2in} 
\frac{\Gamma^2(1+in)}{\pi}\,\tilde{f}e^{-\tau}\,,
\label{OBC}
\ee

\begin{eqnarray}
\Omega^{F}_n(\tau)&\sim&\frac{2\sinh{n\pi}}{\cosh{n\pi}}\tilde{f}\,e^{-\tau}\,,\nonumber\\
C^F_n(\tau)&\sim&2\,i\left(\frac{\tilde{f}}{2}\right)^{-2in}
\frac{\Gamma^2(\frac{1}{2}+in)}{\pi}\tilde{f}\,e^{-\tau}\,. 
\label{OFC}
\end{eqnarray}

\noindent This is the strongly coupled region and the hamiltonian diverge.  
We define the vacuum $|0,p_- \rangle$ as the vacuum seen by asymptotically flat observers at $\tau=\infty$, which is annihilated by the $\alpha_n$ oscillators. We want to define a finite time vacuum ($|0(\tau),p_- \rangle$) such that it is annihilated by time dependent oscillators. The  vacuum $|0(\tau),p_- \rangle$ can be interpreted as the vacuum seen by observers going with the string. We are going to analyze this vacuum as seen by asymptotically flat observers when the string goes towards the singularity. 

\section{The Bogoliubov transformation}

In general, given a pp-wave background with asymptotically flat region at $\tau=+\infty$, we can describe the dynamics of the string  evolving back in time as seen by an observer in the “in” vacuum $|0,p_- \rangle$ at $\tau=\infty$. From the point of view of these observers, there will be string mode creation in the worldsheet vacuum when the string goes towards the singularity at $\tau=-\infty$. Equivalently, one may reverse the orientation of time (which is equal to change the sign of $c$ in $\phi=-cx^+$) and interpret this as an evolution from some excited state to the vacuum at $\tau=+\infty$. 

In this section, we find a unitary Bogoliubov generator  which can be used to construct a finite time Hilbert space from the asymptotically flat  one. This finite time Hilbert space will be related to observers going together with the string towards the singularity. 

 In order to construct the Bogoliubov generators, the following constraints must be taken into account:

a) the Bogoliubov operator must map physical states into physical states;

b) the transformation must be unitary;

c) the Hamiltonian constructed with the new operators must be diagonal as the asymptotic one. 

In the asymptotic limit, a physical closed string state $|\Phi\rangle$  must obey

\begin{equation}
P\left |\Phi\right\rangle= \sum_{n=1}^{\infty }
n\left( N_{n}^{B}+N_{n}^{F}-\bar{N}_{n}^{B}
-\bar{N}_{n}^{F}\right)\left |\Phi\right\rangle=0 \,,
\label{lmc}
\end{equation}

\noindent where $P$ is the momentum and $N^{B}$, $N^{F}$ are the asymptotic boson and fermion number operators. So, the unitary Bogoliubov generator $G$, which generates a new physical closed string Hilbert space from the first one, must satisfy

\begin{equation}
\left[G,\left(N-\widetilde{N}\right)\right]=0 \:.
\label{g}
\end{equation}

\noindent The most general operators (which satisfy the relations (\ref{g})) have the following form 

\begin{equation}
G=G^{B}+G^{F} \:,
\label{gen}
\end{equation}

\noindent for

\begin{equation}
G^{B}=\sum_{n=1} \left(G_{n}^{B} + \bar{G}_n^{B}\right)\:, \,\,\,
G^{F}= \sum_{n=1} \left(G_{n}^{F}
+ \bar{G}_{n}^{F}\right) \:,
\label{genf}
\end{equation}

\noindent where

\begin{eqnarray}
G^{B}_n &=& \lambda _{1_{n}}\widetilde{a}_{n}^{\dagger}
\cdot a^{\dagger }_{n} -\lambda_{2_{n}} a_{n}\cdot \widetilde{a}_{n}
+\lambda_{3_{n}}\left( a^{\dagger }_{n}\cdot a_{n} +
\widetilde{a}_{n} \cdot \widetilde{a}^{\dagger }_{n}\right) \:,
\label{geb} \\
G^{F}_n &=& \gamma_{1_{n}}\widetilde{S}_{n}^{\dagger}
\cdot S^{\dagger }_{n} -\gamma_{2_{n}} S_{n}\cdot \widetilde{S}_{n}
+\gamma_{3_{n}}\left( S^{\dagger }_{n}\cdot S_{n} -
\widetilde{S}_{n} \cdot \widetilde{S}^{\dagger }_{n}\right) \:,
\label{gef}
\end{eqnarray}

\noindent and the transformation parameters will be time dependent. The structure is similar to that of the $SU(1,1)\times SU(2)$ formulation of Thermo Field Dynamics (TFD) developed in \cite{gadcris,gad1,gad2,gad3}, and it is easy to verify that the bosonic generators satisfy the 
$SU\left( 1,1 \right)$ algebra and the fermionic the $SU\left(2\right)$ algebra.

In order to guarantee that the transformation is unitary (constraint b)), we can choose $\lambda_{3_n}$ and $\gamma_{3_n}$ to be real and

\begin{equation}
\lambda_{2_{n}} =-\lambda_{1_{n}}^{*} \,,\,\,\, \gamma_{2_{n}} =-\gamma
_{1_{n}}^{*} \,.
\end{equation}

\noindent For oscillator-like operator $C_n^i$ (which can
be commuting or anti-commuting) and for the asymptotic vacuum $|0,p_-\rangle$, the operators (\ref{geb}) and (\ref{gef}) generate the following unitary transformation:

 \begin{eqnarray}
\left(
\begin{array}{c}
C^{i}_{n}(\Theta) \\
\bar{C}^{i \dagger}_{n}(\Theta)
\end{array}
\right) &=&e^{-iG}\left(
\begin{array}{c}
C^{i}_{n} \\
\widetilde{C}^{i\dagger }_{n}
\end{array}
\right) e^{iG}={\mathbb B}_{n}\left(
\begin{array}{c}
C^{i}_{n} \\
\widetilde{C}^{i \dagger }_{n}
\end{array}
\right) \,, \label{tbt}
\\
\left(
\begin{array}{cc}
C^{i \dagger}_{n}(\Theta) & -\sigma\bar{C}^{i}_{n}(\Theta)
\end{array}
\right) &=&\left(
\begin{array}{cc}
C^{i \dagger }_{n} & -\sigma\widetilde{C}^{i}_{n}
\end{array}
\right) {\mathbb B}^{-1}_{n} \,, 
\label{tbti}\\
|0(\tau), p_- \rangle &=& e^{-iG}|0,p_-\rangle \,,
\label{transf}
\end{eqnarray}

\noindent with $\sigma=1$ for bosons and $\sigma=-1$ for fermions.\footnote{The $\Theta$ represents the parameters of the $G$ transformation.} The operator's matrix transformation is given by

\begin{eqnarray}
{\mathbb B}_{n}=\left(
\begin{array}{cc}
u_{n} & v_{n} \\
\sigma v^{*}_{n} & u^{*}_{n}
\end{array}
\right) \,,\,\,\, |u_{n}|^{2}-\sigma |v_{n}|^{2}=1 \,. 
\label{tbm}
\end{eqnarray}

 \noindent The matrix elements for fermions are

\begin{equation}
u_{n}\equiv U_{n}^{F}=\cosh \left( i\Gamma_{n} \right)
+\frac{\gamma_{3_{n}}}{\Gamma_{n} }
\sinh\left(i\Gamma_{n} \right) \,,
\,\,\,
v_{n}\equiv V_{n}^{F}=-\frac{\gamma _{1_{n}}}{\Gamma_{n} }\mbox{sinh}\left(
i\Gamma_{n} \right) \,, 
\label{uvexpF}
\end{equation}

\noindent and $\Gamma_{n}$ is defined by the following relation

\begin{equation}
\Gamma ^{2}_{n}=-\gamma _{1_{n}}\gamma _{2_{n}}+\gamma
_{3_{n}}^{2} \,.
\label{Gadef}
\end{equation}

\noindent For bosons we have

\begin{equation}
u_{n}\equiv U_{n}^{B}=\cosh \left( i\Lambda_{n} \right) +\frac{\lambda
_{3_{n}}}{\Lambda_{n} } \sinh\left(i\Lambda_{n} \right) \,,
\,\,\,
v_{n}\equiv V_{n}^{B}=\frac{\lambda _{1_{n}}}{\Lambda_{n} }\mbox{sinh}\left(
i\Lambda_{n} \right) , 
\label{uvexpB}
\end{equation}

\noindent and $\Lambda_{n}$ is defined by the following relation

\begin{equation}
\Lambda ^{2}_{n}=\lambda _{1_{n}}\lambda _{2_{n}}+\lambda
_{3_{n}}^{2} \,.
\label{Ladef}
\end{equation}

\noindent So, the explicit form of the state (\ref{transf}) is given by

\begin{eqnarray}
\left |0(\tau), p_-\right\rangle \! &=& \! e^{-i{G}}
\left|0,p_-\right\rangle  \nonumber
\\
&=&
\prod_{n=1}\!\left[\left(
\frac{U_{n}^{F}}{U_{n}^{B}}\right)^{\!\!\!8}\left( \frac{\bar
{U}_{n}^{F}}{\bar{U}_{n}^{B}}\right)^{\!\!\!8}
e^{-\frac{V_{n}^{B}}{U_{n}^{B}}a_{n}^{\dagger}\cdot
\bar{a}_{n}^{\dagger}}
e^{-\frac{V_{n}^{F}}{U_{n}^{F}}
S_{n}^{\dagger}\cdot \bar{S}_{n}^{\dagger}} \right]
\!\!\left|0\right\rangle \,. 
\label{tva}
\end{eqnarray}

Now we just need to fix the parameters using constraint c). Let us start with the bosonic sector. We are going to name the new time dependent bosonic oscillators as $A_n(\tau)$. It is well known from reference \cite{Bin} that the Hamiltonian is diagonalized  by the following transformation for the bosonic fields

\begin{eqnarray}
A^I_n(\tau)&=&\alpha^I_n\,f_n(\tau)+\tilde{\alpha}^I_{-n}\,g^{\ast}_n(\tau) \,,\,\,\,\,
A^{I\dag}_n(\tau)\,\,=\,\,\alpha^I_{-n}\,f^{\ast}_n(\tau)+\tilde{\alpha}^I_{n}\,g_n(\tau)\,,\,\,\,\,\nonumber\\
\tilde{A}^I_n(\tau)&=&\alpha^I_{-n}\,g^{\ast}_n(\tau)+\tilde{\alpha}^I_{n}\,f_n(\tau)\,,\,\,\,\,
\tilde{A}^{I\dag}_n(\tau)\,\,=\,\,\alpha^I_{n}\,g_n(\tau)+\tilde{\alpha}^I_{-n}\,f^{\ast}_n(\tau)\,, 
\label{diagB}
\end{eqnarray}

\noindent where

\begin{eqnarray}
f_n(\tau)&=&\frac{1}{2}\,\sqrt{\frac{\omega_n}{n}}\,\,e^{i\omega_n\tau}\left[Z_n+\frac{i}{\omega_n}\partial_{\tau}Z_n\right] \,,\nonumber\\
g_n(\tau)&=&\frac{1}{2}\,\sqrt{\frac{\omega_n}{n}}\,\,e^{-i\omega_n\tau}\left[-Z_n+\frac{i}{\omega_n}\partial_{\tau}Z_n\right] \,,
\label{fg}
\end{eqnarray}

\noindent and

\begin{eqnarray}
\omega_n=\sqrt{n^2+\tilde{f}^2e^{-2\tau}}\,\,,\,n>0 \,; \,\,\,
\omega_{-n}=-\sqrt{n^2+\tilde{f}^2e^{-2\tau}}\,\,,\,n<0 \:.
\end{eqnarray}

\noindent As

\begin{equation}
\left|f_{n}(\tau)\right|^2 - \left|g_{n}(\tau)\right|^2 =1\,,
\end{equation}

\noindent  we can choose the bosonic Bogoliubov parameters such that
 
\begin{eqnarray}
U_{n}^{B}&=& \cosh \left( i\Lambda_{n} \right) + \frac{\lambda
_{3_{n}}}{\Lambda_{n} } \sinh\left(i\Lambda_{n} \right)= f_n(\tau) \,,\nonumber \\
V_{n}^B&=&\frac{\lambda _{1_{n}}}{\Lambda_{n} }\mbox{sinh}\left(
i\Lambda_{n} \right) = g_n(\tau) \,.
\label{bosonparam}
\end{eqnarray}

\noindent Take notice that the transformation (\ref{diagB}) is exactly the same as (\ref{tbti}). 

Let us deal with the fermions. We are going to write the new time dependent fermionic oscillators as $B_n(\tau)$. From reference \cite{Bin}, the transformation that diagonalizes the fermionic sector is

\begin{eqnarray}
B_n=\cos\varphi_n\,S_n-ie^{i\psi_n}\sin\varphi_n\,\Pi\tilde{S}_n^\dag\,, \,\,\,
B_n^\dag=\cos\varphi_n\,S_n^\dag+ie^{-i\psi_n}\sin\varphi_n\,\Pi\tilde{S}_n\,,\label{B}\\
\tilde{B}_n=\cos\varphi_n\,\tilde{S}_n+ie^{i\psi_n}\sin\varphi_n\,\Pi S_n^\dag\,, \,\,\,
\tilde{B}_n^\dag=\cos\varphi_n\,\tilde{S}_n^\dag-ie^{-i\psi_n}\sin\varphi_n\,\Pi S_n\,, \label{tB}
\end{eqnarray}

\noindent where the constraints that guarantee a diagonal Hamiltonian (written in terms of the B fields) are given by\footnote{See ref. \cite{Bin} for detailed calculations.}

\begin{eqnarray}
e^{i\psi_n}C_n^F=i|C_n^F| \,\,\,\,\, \mbox{and} \,\,\,\,\,
e^{-i\psi_n}C_n^{F\ast}=-i|C_n^F|\,,
\label{psi_n}\\
\sin^2(\varphi_n)= \frac{1}{2}\left[1-\frac{\Omega^F_n}{2\,\omega_n}\right] \,, \label{phi}\\
\tilde{\omega}_n(\tau)=\omega_n(\tau)\equiv\sqrt{n^2+\tilde{f}^2e^{-2\tau}}\,,\label{omega}
\end{eqnarray}

\noindent where $\Omega^{F}_n(\tau)$, $C_n^F$ are given in (\ref{hf}). Note that $B_{-n}=B_n^\dag$ and $\tilde{B}_{-n}=\tilde{B}_n^\dag$ implies that $\varphi_{-n}=-\varphi_n$ and $\psi_{-n}=-\psi_n$. The $\Pi^2=1$ constraint allows us to fix the fermionic Bogoliubov parameters as

\begin{eqnarray}
U_{n}^{F}&=&\cosh \left( i\Gamma_{n} \right)
+\frac{\gamma_{3_{n}}}{\Gamma_{n}}
\sinh\left(i\Gamma_{n} \right)= \cos\varphi_n \,,\nonumber \\
V_{n}^{B}&=&\frac{\lambda _{1_{n}}}{\Lambda_{n}}\mbox{sinh}\left(
i\Lambda_{n} \right)= -ie^{i\psi_n}\sin\varphi_n \,.
\label{ferparam}
\end{eqnarray}

The Bogoliubov transformed Hilbert space is now defined by acting with operators $A_n^ {\dagger}(\tau)$ and $B_n^ {\dagger}(\tau)$ in the vacuum $|0(\tau),p_- \rangle $, which is annihilated by  $A_n(\tau)$ and $B_n(\tau)$, 

\begin{eqnarray}
{A}^{i}_{n}(\tau)|0(\tau),p_- \rangle=
{\tilde A}^{i}_{n}(\tau)|0(\tau), p_- \rangle &=&0 \,,\nonumber \\
{B}^{i}_{n}(\tau)|0(\tau),p_- \rangle=
{\tilde B}^{i}_{n}(\tau)|0(\tau), p_- \rangle &=&0 \,.
\end{eqnarray}

\noindent The vaccum $|0(\tau),p_- \rangle$ is a condensate of $\alpha_n$ and $S_n$ modes and has the structure of a worldsheet left/right entanglement state, as seen by asymptotically flat observers. Owing to property (\ref{g}), this is a condensate of physical states.

The Hamiltonian for the time dependent system is 

\begin{eqnarray}
H(\tau)=H_{F0}+H_{B0}  &+&\frac{1}{\alpha'p^+}\sum^{\infty}_{n=1}\omega_n(\tau)\left[A^I_{-n}(\tau)A^I_n(\tau)
+\tilde{A}^I_{-n}(\tau)\tilde{A}^I_n(\tau)+1\right]\nonumber \\
&+&\frac{1}{\alpha'p^+}\sum_{n=1}^\infty\,
\omega_n(\tau)\left[B_n^\dag(\tau)B_n(\tau)
+\tilde{B}_n^\dag(\tau)\tilde{B}_n(\tau)-1\right].
\end{eqnarray}

\noindent The zero mode is the same as defined in (\ref{H0}) and the frequency is defined in (\ref{omega}).

\section{Two-point function}

In order to obtain more insights of the time dependent superstring sigma model, let us compute the closed string two-point function in this singular background. We will start this study from the bosonic sector. 

Firstly, we need to choose the vacuum. Let's take the vacuum $\left|0,p_-\right\rangle$, which is defined as the Fock space state annihilated by the  $\alpha_n$'s (the flat space vacuum). Using the equations of motion, the two-point function is

\begin{eqnarray}
\left\langle 0, p_{-}|X^I(\sigma,\tau)X^J(\sigma',\tau')|0, p_{-}\right\rangle &=& \delta^{IJ} \frac{\pi}{2}i\alpha'Y_0(\sigma)J_0(\sigma') \nonumber\\ &-&\delta^{IJ}\alpha'\sum_{n=1}^{\infty} \frac{1}{n}Z_n(\tau)Z^*_n(\tau')\cos [2n(\sigma-\sigma')] \,,
\end{eqnarray}

\noindent where the zero mode normal ordering is done in such a way that the momentum is on the right.  In the asymptotic limit $\tau \rightarrow \infty$, $Z_n(\tau) \sim e^{2ni \tau}$,  $Y_0(\sigma)J_0(\sigma')\sim \frac{2}{\pi}(\ln {\tilde f}- \tau)$, and the two-point function corresponds to the usual flat space result.

Let's focus on non zero modes. For simplicity, we will evaluate the two-point function at equal times. We will compute the function

\begin{equation}
F(\sigma,\sigma',\tau)= \sum_{n=1}^{\infty} \frac{1}{n}Z_n(\tau)Z^*_n(\tau)\cos [2n(\sigma-\sigma')]= \sum_{n=1}^{\infty} \frac{|J_{in}(u)|^2}{\sinh(n\pi)}\cos[2n(\sigma-\sigma')] \,,
\label{Fosc}
\end{equation}

\noindent where we have used the equations (\ref{Tn}) and (\ref{Z}), and $u={\tilde f}e^{-\tau}$. For the case of the background (\ref{btsey}), in \cite{Ryang} the two-point function was calculated using the approximation $n\tau<<1$  and going to the continuous limit. Here this trick is not possible owing to the term $\sinh(n\pi)$ that appears in (\ref{Fosc}).  We are going to take a different approach. In order to study the behavior of the two-point function close to flat space, we use the power series

\be
J_{\mu}(u)J_{\nu}(u)=\frac{1}{2}u^{\mu+\nu}\sum_{k=0}^{\infty}\frac{(\mu+\nu+k+1)_k(\frac{1}{4}u^2)^k}{k!\Gamma(\mu+k+1)\Gamma(\nu+k+1)} \,,
\label{prodE}
\ee

\noindent where $(a)_n = \Gamma (a+n) / \Gamma (a)$ is the Pochhammer's symbol. We have an expression for each order of $u = {\tilde f}e^{-\tau}$.  At zero order, we get the flat space result

\begin{eqnarray}
|J_{in}(u)|^2&=&  \frac{1}{\pi^2}\cosh(\pi n)\frac{|\Gamma(1/2+in)|^2}{|\Gamma(i+in)|^2} =\frac{1}{\pi n}\sinh(\pi n) \,, \nonumber \\
F(\sigma,\sigma',\tau)&=& \sum_{n=1}^{\infty} \frac{1}{n}\cos [2n(\sigma-\sigma')] = \ln [2\sin(\sigma-\sigma')] \,.
\end{eqnarray}

\noindent For the first correction in $u$, equation (\ref{prodE}) gives

 \begin{equation}
|J_{in}(u)|^2= \frac{u}{\pi^2}\cosh(\pi n)\frac{|\Gamma(1/2+in)|^2}{|\Gamma(i+in)|^2}\left(1+\frac{a}{n^2+1}\right) \,,
\end{equation}

\noindent where we have employed the following properties of the Gamma function: $\Gamma(x+1)=x\Gamma(x)$ and $a = \Gamma^2(3/2)+2\Gamma(3/2)\Gamma(1/2)$. At this order, the function (\ref{Fosc}) is

\begin{equation}
F(\sigma,\sigma',\tau)= (1+a)\sum_{n=1}^{\infty} \frac{1}{n}\cos [2n(\sigma-\sigma')] -a\sum_{n=1}^{\infty} \frac{n}{n^2+1}\cos [2n(\sigma-\sigma')] \,.
\label{fo1}
\end{equation}

\noindent In (\ref{fo1}) the first term is identical to the zero-order term. Let's focus on the second sum, which can be written in terms of Hypergeometric functions,

\begin{eqnarray}
\sum_{n=1}^{\infty} \frac{n}{n^2+1}\cos [2n(\sigma-\sigma')]=\Re\frac{e^{\frac{i\pi}{4}}}{2\sqrt{2}}e^{i(\sigma-\sigma')} [ {}_2F_1(1 + i, 2, 2 + i, e^{i (\sigma-\sigma')}) \nonumber\\
 -i\:{}_2F_1(1-i, 2, 2 -i, e^{i(\sigma-\sigma')})] \,.
\end{eqnarray}

\noindent Now we will show that, at this order, the behavior at short distances of the two-point function is the same as that of the flat space. The Hypergeometric function ${}_2F_1(a,b,c;z)$ is analytic everywhere except for possible branch points at $z=0,1$ and $\infty$; we are going to work with the principal branch, and use the usual linear transformation of the Hypergeometric functions:

\be
{}_2 F_1 (a,b,c;z)= (1-z)^{c-a-b} {}_2 F_1 (a,c-a,c-b,c;z) \,.
\label{TF}
\ee

\noindent As shown in \cite{watson}, ${}_2 F_1 (a,b,c;z)$ is regular at $z=1$ if $\Re(c-a-b)> 0 $; in this case,

\be
{}_2 F_1 (a,b,c;1)=\frac{\Gamma(a)\Gamma(c-a-b)}{\Gamma(c-a)\Gamma(c-b)} \,.
\ee

\noindent  Using this result and the transformation (\ref{TF}), the short-distance behavior of the Hypergeometric functions

\be
{}_2 F_1(1-i,2,2-i;z\rightarrow 1) &=&\frac{\Gamma(2-i)}{\Gamma(1-i)}\left(\frac{1}{1-z}\right) \,,\nonumber \\
{}_2 F_1(1+i,2,2+i,z\rightarrow 1)&=&\frac{\Gamma(2+i)}{\Gamma(1+i)}\left(\frac{1}{1-z}\right) \,,
\ee

\noindent cancels each other in (\ref{fo1}) and the only singular behavior comes from the first term of (\ref{fo1}); so, the two-point function has the same short-distance behavior as the flat space string. 

Close to the background singularity, using the asymptotic expansion (\ref{ase}), the equal times bosonic two-point function is

\begin{equation}
\lim_{\tau \rightarrow -\infty} \left\langle 0, p_-|X^I(\sigma,\tau)X^J(\sigma',\tau)|0, p_-\right\rangle = - \delta^{IJ} \frac{\alpha'}{ u}\sum_{n=1}^{\infty} \frac{|\cos(u'+\frac{in\pi}{2})|^2}{\sinh(n\pi)}=- \delta^{IJ} \frac{\alpha'}{ u}G(\sigma,\tau) \,, 
\label{2psing}
\end{equation}

\noindent where $u'=u-\pi/4$, and

\be
G(\sigma,\tau)= \sum_{n=1}^{\infty} \frac{\cos(2u')\cos [n(\sigma-\sigma')]}{\sinh(n\pi)}+  \sum_{n=1}^{\infty} \frac{\cosh(n\pi)\cos [n(\sigma-\sigma')]}{\sinh(n\pi)} \,.
\label{sH}
\ee

\noindent The sum can be written as a q-Lambert series:

\be
\sum_{n=1}^{\infty} \frac{w^n}{1-q^{2n}}=\frac{1}{2\ln(q)}\left(\ln(1-q^2)-\Psi_{q^2}\left(\frac{\ln w}{\ln q^2}\right)\right) \,,
\label{s2}
\ee

\noindent where $\Psi_q$ is the q-Pollygamma function. The q-Gamma and q-Polygamma functions are defined  from the q-factorial, $(a;q)_\infty=\displaystyle\prod_{k=1}^{\infty}(1-aq^k),|q|<1$:

\be
\Psi_q(z)&=&\frac{d}{dz}\ln\Gamma_q(z) \,, \nonumber \\
\Gamma_q(z)&=& \frac{(q;q)_\infty(1-q)^{1-z}}{(q^z;z)_\infty} \,.
\ee

\noindent For $q=e^{-\pi}$, $z=i(\sigma-\sigma')$, the sum in (\ref{sH}) is

\be
G (\sigma, \tau)&=&\frac{1}{2\ln(q)}\ln(1-q^2)\nonumber \\
&-&\frac{1}{8\ln(q)}[\Psi_{q^2}(iz-\pi)+\Psi_{q^2}(-iz-\pi)+\Psi_{q^2}(iz-2\pi)+\Psi_{q^2}(-iz-2\pi)+\Psi_{q^2}(iz) \nonumber\\
&+&\Psi_{q^2}(-iz)] \,.
\label{spsi}
\ee

\noindent Using the relation

\be
\Gamma_q(z) =\frac{1-q}{1-q^z}\Gamma_q(z+1) \,,
\ee

\noindent we can easily see that, when $z$ goes to zero, the q-Polygamma function $\Psi_{q}(z)$ has a singularity like $1/z$. The only contribution to the two-point function short-distance  singularity comes from the last two terms in (\ref{spsi}) and they cancel each other out. So, in the vacuum $\left|0,p_-\right\rangle$, the short-distance behavior of the two-point function close to background singularity is not singular. 

Now, we turn our attention to the fermionic two-point function in the vacuum $|0,p_-\rangle$. The fermionic two-point function has problems in the flat asymptotic limit if we start from the expansion in modes used in equation (\ref{mode})

\be
\left\langle 0, p_-|\theta^1(\sigma,\tau)\theta^1(\sigma',\tau)|0, p_-\right\rangle=\frac{1}{4\pi \xi}\sum_{n=1}^{\infty} (|W(\tau)|^2e^{in(\sigma-\sigma')}+ |\tilde{W}(\tau)|^2e^{-in(\sigma-\sigma')}
) \,.
\label{2ptf1}
\ee
 
\noindent Let's analyze the fermionic two-point function close to the flat space limit.  Expanding the Gamma functions using (\ref{prodE}), we have a power series  in $u^2 = {\tilde f}^2e^{-2\tau} $. Up to first order

\be
\left\langle 0, p_-|\theta^1(\sigma,\tau)\theta^1(\sigma',\tau)|0, p_-\right\rangle&=& \frac{1}{4\pi \xi}\left[\sum_{n=1}^{\infty} e^{in(\sigma-\sigma')}+\frac{u^2}{4}\sum_{n=1}^{\infty} \frac{|\Gamma(1/2+in)|^2\cos [n(\sigma-\sigma')]}{|\Gamma(3/2+in)|^2} \right] + {\mathcal O} (u^4) \nonumber \\
&=& \frac{1}{4\pi \xi}\left[\sum_{n=1}^{\infty} e^{in(\sigma-\sigma')}+\frac{u^2}{4}\sum_{n=1}^{\infty} \frac{\cos [n(\sigma-\sigma')]}{n^2+1/4} \right ] + {\mathcal O} (u^4) \,.
\label{2ptf}
\ee

\noindent The time dependent term can be evaluated by residue theorem; the result is

\be
\sum_{n=1}^{\infty} \frac{\cos [n(\sigma-\sigma')]}{(n^2+1/4)} = \frac{\pi \cosh \left(\frac{\pi-|\sigma-\sigma'|}{2}\right)}{\sinh(\pi/2)}-2 \,.
\ee

\noindent Up to this order, there is no short distance singularity in the fermionic two-point function. The only singular contribution comes from the time independent term in (\ref{2ptf}), that is, the flat space contribution. Now we will show that this behavior remains at the asymptotic limit $\tau \rightarrow -\infty$. Using the asymptotic limit of the Bessel function given in (\ref{ase}),  we have

\be
|W(\tau)|^2&\rightarrow& \frac{1}{2}\left[\frac{\cos 2u}{\cosh n\pi}+1\right] \,, \nonumber \\
|\tilde{W}(\tau)|^2&\rightarrow& \frac{1}{2}\left[-\frac{\cos 2u}{\cosh n\pi}+1\right] .
\ee

\noindent In this limit the fermionic two-point function can be written as

\be
\lim_{\tau\rightarrow\infty}\left[\left\langle 0, p_-|\theta^1(\sigma,\tau)\theta^1(\sigma',\tau)|0, p_-\right\rangle\right]
=\sum_{n=1}^{\infty}\cos n(\sigma-\sigma') + \cos(2u) \sum_{n=1}^{\infty}\frac{v^n-\bar{v}^n}{1+q^{2n}} \,,
\ee

\noindent where $v= e^{-\pi}e^{in(\sigma-\sigma')}$  and $q$ is the same one we have used before. The first sum is identical to the flat space result; the second one can be written as the followig q-Lambert series:

\be
\sum_{n=1}^{\infty} \frac{w^n}{1+q^{2n}}= \frac{1}{2\ln q}\left(\ln(1+q^2)-\Psi_{q^2} \left(\frac{\ln w}{2\ln q}\right)+ \Psi_{q^4} \left(\frac{\ln w}{4\ln q}\right)\right) \,.
\ee

\noindent So, as $\tau\rightarrow -\infty$, with $z=i(\sigma-\sigma')$ and $q=e^{-\pi}$, we get

\be
\lim_{\tau\rightarrow -\infty}\left[\left\langle 0, p_-|\theta^1(\sigma,\tau)\theta^1(\sigma',\tau)|0, p_-\right\rangle\right]
= \sum_{n=1}^{\infty} \cos [n(\sigma-\sigma')] \nonumber \\
+\frac{1}{2\pi}\cos(2u)\left(\Psi_{q^2}\left(\frac{\pi -z}{2\pi}\right)-\Psi_{q^2}\left(\frac{\pi - \bar{z}}{2\pi}\right)\right) \nonumber \\
+\frac{1}{2\pi}\cos(2u)\left(\Psi_{q^4}\left(\frac{\pi -\bar{z}}{4\pi}\right)- \Psi_{q^4}\left(\frac{\pi -z}{4\pi}\right)\right) \,. 
\ee

\noindent Once more, the only singular term as $z\rightarrow 0$ is like the flat space one (the first term).  

Henceforth, we turn our attention to the Bogoliubov transformed vacuum $|0,\tau \rangle=e^{iG}\left|0,p_-\right\rangle$. Using the inverse of the Bogoliubov transformation defined in (\ref{diagB}), the bosonic two-point function in this vacuum is

\be
\left\langle 0, p_-|\exp(-iG)X^I(\sigma,\tau)X^J(\sigma',\tau')\exp(iG)|0, p_-\right\rangle&=& \delta^{IJ}\frac{\pi}{2}i\alpha'Y_0(\sigma)J_0(\sigma')+ \delta^{IJ} \alpha' \sum_{n=1}^{\infty}\frac{1}{\omega_n}\cos [n(\sigma-\sigma')]\nonumber \\
&=&\delta^{IJ} \frac{\pi}{2}i\alpha'Y_0(\sigma)J_0(\sigma')+ \delta^{IJ} \alpha'\sum_{n=-\infty}^{\infty} \frac{1}{\omega_n}e^{in(\sigma-\sigma')}. \nonumber \\
\ee

\noindent In this case, the strategy is the same used in \cite{Ryang}. The mode summation is performed through the Poisson resummation formula:

\begin{eqnarray}
\sum_{n=-\infty}^{\infty} \frac{1}{\omega_n}e^{in(\sigma-\sigma')}
&=& \sum_{l=-\infty}^{\infty}\int_{-\infty}^{\infty}dx 
e^{2 \pi ixl}\frac{1}{\sqrt{x^2+m^2}}e^{ix(\sigma - \sigma')} \nonumber \\
&=& 2K_0(m(\tau)|\sigma-\sigma'|) + 2\sum_{l \neq 0} K_0(m(\tau)|l\pi + \sigma-\sigma'|) \,,
\label{ko}
\end{eqnarray} 

\noindent with $m(\tau)={\tilde f}e^{-\tau}$ and $K_0(x)$ is the modified Bessel function. Unlike the $\left|0,p_-\right\rangle$ vacuum, in the transformed Bogoliubov vacuum we do not have the separation between the temporal and the ($\sigma-\sigma'$) dependence . We have two different behaviors separated by the similar point ${\tilde f}|\sigma-\sigma'| \sim e^{\tau}$.  For $\frac{{\tilde f}|\sigma-\sigma'|}{e^{\tau}}<<1$, we  can use

\begin{equation}
K_0(x) = -I_0(x)\ln\left( \frac{x}{2}\right) + \sum_{k=0}^{\infty}
\frac{\psi(k+1)}{(k!)^2} \left(\frac{x}{2}\right)^{2k} \,,
\label{kol}
\end{equation}

\noindent to show that the leading short-distance behavior of  the two-point function is

\begin{equation}
\delta^{IJ}\alpha' \ln (m(\tau)|\sigma-\sigma'|) \,.
\label{shr}\end{equation}

In the region $\frac{{\tilde f}|\sigma-\sigma'|}{e^{\tau}}>>1$ we have a totally different behavior. Using asymptotic expansion of the modified Bessel function, 

\be
K_0(x)= \sqrt{\frac{\pi}{2x}}e^{-x}\sum_{k=0}^{\infty}\frac{\Gamma(k+1/2)}{k!\Gamma(1/2-k)}(2x)^{-k} \,,
\ee

\noindent the leading order term comes from the first term in (\ref{ko}) and it is given by 

\begin{equation}
\delta^{IJ}\alpha' \sqrt{\frac{e^{\tau}}{2{\tilde f}|\sigma-\sigma'|}}
e^{-\frac{{\tilde f}|\sigma-\sigma'|}{e^{\tau}}}
\sum_{n=0}^{\infty}\frac{\Gamma(n+\frac{1}{2})}
{n!\Gamma(-n+\frac{1}{2})}\left(\frac{e^{\tau}}{{\tilde f}|\sigma-\sigma'|}\right)^n .
\label{epp}
\end{equation}

\noindent Now we have an exponential damping which resembles the exponential tail of the 
positive energy part of the invariant $\Delta$ function for the massive scalar field in the space-like separation; also, the result given in (\ref{ko}) resembles a real time finite temperature two-point function, with temperature $T=\frac{1}{2\pi}$ \cite{KimLee}. Actually, due to the Bogoliubov transformation, there is a correlation between $X_R$ and $X_L$; it will be shown in Section 6 that this correlation is related to the thermalization of the system. 

For the fermionic sector, the two-point function calculated in the Bogoliubov transformed vacuum is calculated using the inverse of the Bogoliubov transfomations

\be
e^{iG}\beta_n e^{-iG}= B_n(\tau)\cos\varphi_n+ie^{\psi_n}\sin\varphi_n\Pi\tilde{B}^{\dagger}(\tau) \,, \nonumber \\
e^{iG}\tilde{\beta}_n e^{-iG}= \tilde{B}_n(\tau)\cos\varphi_n-ie^{\psi_n}\sin\varphi_n\Pi{B}^{\dagger}(\tau) \,.
\ee

\noindent The fermionic two-point function in the Bogoliubov transformed vacuum is 

\be
\langle 0,\tau|\theta(\sigma,\tau)\theta(\sigma',\tau)|0,\tau\rangle= \langle 0, p_-|\theta(\sigma,\tau)\theta(\sigma',\tau)|0,p_-\rangle \nonumber \\
+ \frac{i}{2\pi \xi} \sum_{n=1}^{\infty}(|\tilde{W}(\tau)|^2-|W(\tau)|^2)\sin [n(\sigma-\sigma')]\sin^2\varphi_n \,, \label{ferm2pt}
\ee

\noindent where the first line in the equation is the two-point function in the time independent  vacuum. At limit $t\rightarrow\infty$, we get $\sin^2\varphi_n \rightarrow 0$ and we have $\langle 0,\tau|\theta(\sigma,\tau)\theta(\sigma',\tau)|0,\tau\rangle= \langle 0, p_-|\theta(\sigma,\tau)\theta(\sigma',\tau)|0,p_-\rangle$ as expected. Close to singularity, the function $\sin^2\varphi_n$ (using(\ref{phi})) has the form of a Fermi-Dirac distribution 

\be
\lim_{\tau\rightarrow-\infty}\sin^2\varphi_n= \frac{1}{e^{2\pi n}+1}\,. \label{fdirac} 
\ee
The second term in (\ref{ferm2pt})  can be written as q-Polygamma functions

\be
\lim_{\tau\rightarrow -\infty}\sum_{n=1}^{\infty}(|\tilde{W}(\tau)|^2-|W(\tau)|^2)\sin [n(\sigma-\sigma')]\sin^2\varphi_n
= - \frac{\cos 2u}{2\pi} \sum_{l=0}^{\infty} \left[\Psi_{q^2} \left(\frac{z' +2\pi l}{2\pi}\right) \right. \nonumber\\
\left. - \Psi_{q^2} \left(\frac{{\bar z'} +2\pi l}{2\pi}\right) + \Psi_{q^4} \left(\frac{{\bar z'} +2\pi l}{4\pi}\right) - \Psi_{q^4} \left(\frac{z' +2\pi l}{4\pi}\right)\right] \,. \nonumber\\
\ee

\noindent where $z' = 3\pi -i(\sigma - \sigma')$. Again, we do not have any new singular term in the short distance behavior. As the bosonic case, the fermionic one resembles a real time finite temperature two-point function. 

\section{The left/right entropy production}

Let us define the density matrix

\begin{equation}
\rho(\tau)= |0(\tau), p_-\rangle\langle 0(\tau), p_-| \,,
\end{equation}

\noindent where the state $|0(\tau), p_-\rangle$ is given in (\ref{tva}) with definitions (\ref{bosonparam}) and (\ref{ferparam}). The R reduced density matrix is calculated  by tracing over the left degrees of freedom. In order to make clear the calculation we intend to do, we are going to fix the following notation: the $I$ index in the bosonic $|n_k^I\rangle_B$ and ferminonic state $|n_k^I\rangle_F$  ranges from 1 to $d$, where $d=8$; a state without index $B$ or $F$ means the tensor product of the fermionic and bosonic states; the sum $\displaystyle\sum_{n_k}$ means $\displaystyle\sum_{n_1,n_2,...}$; and the sum $\displaystyle\sum_{\{M_k\}}$ means $\displaystyle\sum_{n_k,m_k,o_k,p_k}$. Keeping in mind this notation, the R reduced density matrix is

\begin{eqnarray}
\rho_R(\tau)=Tr_L|0(\tau), p_-\rangle\langle 0(\tau), p_-| \nonumber\\
=\prod_{k=1}^{\infty}|f_k(\tau)\cos\varphi_k|^{-16}\prod_{I=1}^{8}\sum_{\{M_k\}}^{\infty}\sum_{l_k=0}^{\infty}\(\frac{g^\ast_k(\tau)}{f_k(\tau)}\)^{n_k}\(\frac{g_k(\tau)}{f^\ast_k(\tau)}\)^{m_k}\langle l_k^{I}|n_k^{I}\rangle_B \ \langle m_k^{I}|l_k^{I}\rangle_B \nonumber \\
 \times\(\tan\varphi\)^{o_k}\(\tan\varphi\)^{p_k}\langle l_k^{I}|o_k^{I}\rangle_F  \langle p_k^{I}|l_k^{I}\rangle_F |n_k^{I}\rangle_B\langle m_k^{I}|_B |o_k^{I}\rangle_F \langle p_k^{I}|_F\nonumber\\
=\prod_{k=1}^{\infty}|f_k(\tau)\cos\varphi_k|^{-16}\prod_{I=1}^{8}\sum_{n_k,m_k}\left|\frac{g_k(\tau)}{f_k(\tau)}\right|^{2n_k}\tan^{2n_k}\varphi_k|n_k^{I}\rangle_B \langle n_k^{I}|_B |m_k^{I}\rangle_F \langle m_k^{I}|_F \,.
\label{dens}
\end{eqnarray}

\noindent A measure of the bosonic left/right entanglement of the state $ |0(\tau), p_- \rangle$ is given by the von Neumann entropy associated with the R reduced density matrix:

\begin{eqnarray}
S &=&-Tr\rho_{R}\ln\rho_{R} \nonumber \\
  &=&  -8\sum_{n=1}^{\infty}[|g_{n}(\tau)|^2\ln ( |g_{n}(\tau)|^2)
-(1+|g_{n}(\tau)|^2 )\ln (1 +|g_{n}(\tau)|^2 )] \nonumber \\
&-&  8\sum_{n=1}^{\infty}[\sin^2\varphi_n\ln ( \sin^2\varphi_n)
+\cos^2\varphi_n\ln (\cos^2\varphi_n)] \,,
\label{s}
\end{eqnarray}

\noindent where $|g(t)|^2$ can be written as

\begin{eqnarray}
|g_{n}(\tau)|^2&=&\frac{1}{2}\left[\frac{\Omega^B_n}{2\,\omega_n}-1\right] \,,
\label{NB}
\end{eqnarray}

\noindent and $\sin^2 \varphi_n$ is defined in (\ref{phi}). Please note that the entanglement is produced by the background and it is seen by an observer in the vacuum $|0,p_- \rangle$ at $\tau=\infty$. If we take the time derivative of $S$, we get

\begin{eqnarray}
\dot{S}&=&-8\displaystyle{\sum_k}\dot{N^B_k}(\tau)\ln\left(\frac{1+N^B_k(\tau)}{N^F_k(\tau)}\right) \nonumber \\
&-&8\displaystyle{\sum_k}\dot{N^F_k}(\tau)\ln\left(\frac{1+N^F_k(\tau)}{N^F_k(\tau)}\right) \,,
\end{eqnarray}

\noindent where $N^B_k(\tau)= |g(\tau)|^2$ and $N^F_k=\sin^2\varphi_k$. So, the signal of $\dot{S}$ can be read just from $\dot{N^B_k}(\tau)$ and $\dot{N^F_k}(\tau)$.

The functions $N^B_k(\tau)$ and $N^F_k(\tau)$ are related to the expected value of the number operator of the original Hilbert space

\begin{eqnarray}
\langle 0(\tau), p_-|\alpha_k^{\dagger}\alpha_k | 0(\tau), p_-\rangle &=& 8 |g(\tau)|^2 \,, \nonumber\\
\langle 0(\tau), p_-|S_k^{\dagger}S_k | 0(\tau), p_-\rangle &=& 8 |g(\tau)|^2 \,. \label{NO}
\end{eqnarray} 

\noindent Using the following properties of Bessel functions 

\begin{eqnarray}
\dot{J}_{\nu} (z) = \frac{1}{2} [J_{\nu -1} (z) + J_{\nu + 1} (z)] \: , \nonumber\\
\frac{2\nu}{z} J_{\nu} (z) = J_{\nu -1}(z) + J_{\nu +1}(z) \:,  \nonumber \\
J_{-\frac{1}{2} + in} (z) J_{-\frac{1}{2} - in} (z) + J_{\frac{1}{2} + in} (z)J_{\frac{1}{2} - in} (z) = \frac{2 \cosh n\pi}{\pi z} \:,
\end{eqnarray}

\noindent and the expressions

\begin{eqnarray}
\Omega^F_n &=& \frac{2i\pi u}{\cosh n\pi} \left[ \frac{\cosh n\pi}{2\pi u} + G(u) \right] \:, \\
G(u) &=& \frac{u}{2} [J_{-\frac{1}{2} +in} (u) J^{'}_{-\frac{1}{2} -in} (u) + J_{\frac{1}{2} + in} (u) J^{'}_{\frac{1}{2}-in} (u)] \:, \\
\end{eqnarray}

\noindent together with the Bessel function's equations, we get 

\begin{eqnarray}
\dot{N}^B_{n} (\tau) &=& \frac{\tilde{f}^2 e^{-2\tau}}{4n\omega^3_n} \left[ |\partial_{\tau}Z_n|^2 \left (1 - \frac{\tilde{f}^4 e^{-4\tau}}{n^2} \right) - n^2 |Z_n|^2  \right] \,, \nonumber \\
\dot{N}^F_{n}(\tau) &=& -\frac{\pi n \tilde{f} e^{-\tau}}{4\omega^3_n \cosh n\pi} \left[2n \mbox{Im}\: G + \omega^2_n (|J_{-\frac{1}{2} + in}|^2 + |J_{\frac{1}{2} +in}|^2 )\right] \:.
\end{eqnarray}

\noindent One can see that, close to the singularity ( $\tau\rightarrow -\infty$), the dominant terms in the equations above are negative and
 the entropy increases with time.\footnote{In \cite{Bin} it is proven that Im $G$ is negative definite, but this is not the dominant term for $\tau \rightarrow -\infty$.} We are going to show that close to the singularity the worldsheet thermalizes; but before it, let us discuss the condition for a maximum entanglement of the worldsheet left/right movers at a fixed time. This is the condition for the state (\ref{tva}) to be a maximum entanglement state; in other words, $S$ to be maximum. We see that $S$ is a function of the expected value of the right number operator; that is, $S$ is a function of the expected value of the right Hamiltonian on the state (\ref{tva}). Based on that, let us define the constraints

\begin{equation}
E_R= \langle 0(\tau)| H_R|0(\tau)\rangle= Tr\rho_R H_R ,\: \: \: Tr\rho_R=1 \,, \label{const}
\end{equation}  

\noindent where $H_R$ is the right mover Hamiltonian for observers in the
vacuum $|0,p_- \rangle$ at $\tau=\infty$.  We can show the maximum entanglement condition in a ``thermodynamical way". The maximal left/right entanglement is achieved demanding that $\delta S(E_R)=0$ under the constraint (\ref{const}). So, the state (\ref{tva}) is a maximum entanglement state at a fixed time if the density matrix (\ref{dens}) can be written as a Gibbs-like density operator

\begin{eqnarray}
\rho_R&=& \frac{1}{Z_R}e^{\gamma H_R} \,,\nonumber \\
Z_R&=& Tr e^{\gamma H_R} \,,
\label{mec}
\end{eqnarray}

\noindent for some parameter $\gamma$ defined at each time; this is just the case of the state (\ref{tva}). The properties (\ref{tbm}) allow us to define

\begin{eqnarray}
|F_k(\tau)|^2&=& \frac{1}{1-e^{-\gamma k}} \,, \nonumber \\
|\cos\varphi_k|^2 &=& \frac{1}{1+e^{-\gamma k}} \,,
\end{eqnarray} 

\noindent so

\begin{eqnarray}
|g_k(\tau)|^2&=& \frac{1}{e^{\gamma k}-1} \,, \nonumber \\
|\sin\varphi_k|^2&=& \frac{1}{1+e^{\gamma k}} \,, \nonumber \\
\left|\frac{g_k(\tau)}{F_k(\tau)}\right|^2&=&\left|\frac{\sin^\varphi_k}{\cos\varphi_k}\right|=e^{-\gamma k} \,, \nonumber \\
Z_R&=& \prod_{k=1}^{\infty}\frac{1}{1-e^{-2\gamma k}} = \prod_{k=1}^{\infty}|F_k(\tau)\cos\varphi_k|^2 \,.
\end{eqnarray}

\noindent With these definitions, the density matrix can be written as

\begin{eqnarray}
\rho_A(t)&=&Tr_B|0(\tau),p_-\rangle\langle 0(\tau),p_-| \nonumber\\
&=&\prod_{k=1}^{\infty}\frac{1}{1-e^{-2\gamma k}}\sum_{n_k,m_k}\left[\tanh(\gamma)\tanh(\bar{\gamma})\right]^{n_k}\left[\tanh(\gamma)\tanh(\bar{\gamma})\right]^{m_k}|n_k\rangle\langle n_k|_B |n_k\rangle\langle n_k|_F \,, \nonumber\\
\end{eqnarray}

\noindent which has the form of (\ref{mec}).\footnote{Note that $e^{\gamma H(t)}$ is not truly a density operator from the Liouville-von Neumann (LvN) equation. However, for a fixed time, $S$ is an increasing function of energy.}

\subsection{Entropy Operator}
 
We will now show that the entanglement entropy calculated earlier has the same form  as a thermodynamic entropy. The inspiration for this is the Thermo Field Dynamics (TFD) \cite{ume1, ume2}, a canonical finite temperature formalism.\footnote{For applications of TFD in superstring theory see, for example,  \cite{adsnos, gad3, nedel1, torus, nedel2}.} The central idea of the TFD formalism is the doubling of degrees of freedom and a Bogoliubov transformation to entangle such duplicated degrees, defining a thermal vacuum. The temperature is introduced as an external parameter, and the thermal vacuum appears as a boundary state in the doubled Fock space composed by the physical space of the system and a copy of it.  The structure of the time dependent left/right entanglement state found in this work is very similar to the TFD thermal vacuum.

 In the TFD formalism, there is an operator whose expected value in the thermal vacuum provides the thermodynamic entropy of the system. Inspired by the TFD entropy operator, let's define the following time-dependent operator:

\be
K(\tau)= &-&\sum_{n=1}^{\infty}\left[\alpha_{n}^{\dagger}\cdot \alpha_{n}
\ln\left(|g_{n}(\tau)|^{2}\right)-\alpha_{n}\cdot
\alpha_{n}^{\dagger}\ln\left(|f_{n}(\tau)|^{2}\right)\right] \nonumber\\
&-&\sum_{n=1}^{\infty}\left[S_{n}^{\dagger}\cdot {S}_{n}
\ln\left(|\sin\varphi_n|^{2}\right)-S_{n}\cdot
{S}_{n}^{\dagger}\ln\left(|\cos\varphi_n|^{2}\right)\right] \,.
\ee

\noindent If we take the expected value of the operator $K$ in the vacuum $ |0(\tau), p_- \rangle $ using the relations (\ref{NO}), we get exactly the entanglement entropy calculated in (\ref{s}). This operator has another important feature: it can be used as an entanglement state generator. It can be shown that

\be
|0(\tau),p_-\rangle =e^{-K(\tau)}e^{\sum \alpha_{n}^{\dagger}\cdot \widetilde{\alpha}_{n}^{\dagger}}e^{\sum S_{n}^{\dagger}\cdot \widetilde{S}_{n}^{\dagger}}|0,p_-\rangle \,.
\ee

\noindent This expression brings to light an important characteristic concerning the entanglement dynamics. Note that the entropy operator carries all the temporal dependence of the state; in particular, we can verify that the time evolution of the entanglement state's vacuum is generated
by the time derivative of the entropy operator:

\be
\frac{\partial\left|0(\tau),p_-\right\rangle}{\partial \tau}=-\frac{1}{2}\frac{\partial K}{\partial \tau}\left|0(\tau),p_-\right\rangle \,. 
\label{eq-mov-entropy}
\ee

\noindent This equation implies that the basic notion of equilibrium, $\displaystyle\frac{\partial\left|0(\tau)\right\rangle}{\partial \tau}\approx0$, is equivalent to the the maximum entropy condition. This condition is achieved when the string approaches the singularity and the entropy becomes the thermodynamic entropy, as we will show in the next section.  
If we calculate the following projection

\begin{equation}
\langle p_-, 0 \mid 0(\tau),p_-\rangle= e^{-8
\sum_{n}\ln\left(1+|g_{n}(\tau)|^{2}\right)} e^{-8\sum_n \ln(|\cos\varphi_n |^2)} \,,
\label{ket}
\end{equation}

\noindent and take the limit $\tau \rightarrow -\infty$ , we get 
$\langle p_-, 0 \mid 0(\tau),p_- \rangle = 0$, showing that the state close to the singularity  is unitarily inequivalent to the asymptotically flat vacuum. In other words, this result shows that close to the singularity the system defined by the 2d worldsheet quantum field theory is led to another representation of the canonical commutation relations, which is unitarily inequivalent to the representation at $\tau =\infty$.  This is typical of entanglement states, as show in \cite{BEY}, but also is a general characteristic of quantum dissipative theories \cite{garvit},\cite{ceravi} and  thermal theories \cite{ume2, ume1}. In all these scenarios the non unitary evolution  seems to be generated by the same kind of entropy operator.

\section{Thermalization}

In this section, we are going to analyze the entanglement state near the singularity. The time dependent left/right entanglement state is 

\begin{eqnarray}
\left |0(\tau), p_-\right\rangle
=
\prod_{n=1}^{\infty} \!\left[\left(
\frac{\cos^2\varphi_n}{|f_{n}(\tau)|^2}\right)^{\!\!\!8}
e^{-\frac{g_{n}(\tau)^{*}}{f_{n}(\tau)}\a_{n}^{\dagger}\cdot
\bar{\a}_{n}^{\dagger}}
e^{-{\Pi e^{i\psi_n}\tan\varphi_n}
S_{n}^{\dagger}\cdot \bar{S}_{n}^{\dagger}} \right]
\!\!\left|0,p_-\right\rangle \,.
\label{estfg}
\end{eqnarray}

\noindent For the bosonic part of the state, we need the expression

\be
\lim_{\tau\rightarrow -\infty}\left[\frac{g^*_n(\tau)}{f_n (\tau)}\right ]= -\frac{e^{-2in\ln(f/2)}\Gamma(1+in)}{\Gamma(1-in)}\lim_{u\rightarrow \infty}\frac{J_{-in}(u)-iJ_{-in}'(u)}{J_{in}(u)-iJ'_{in}(u)} \,.
\ee

\noindent Using
 
\be
\Gamma (1 \pm in)&=& \frac{n\pi}{\sinh n\pi}e^{\pm i\gamma_n} \,, \nonumber\\
\gamma_n &=& n\psi(x) +\sum_{k=0}^{\infty}\left[\frac{n}{1+k}-\arctan \frac{n}{1+k}\right ] \,,
\ee

\noindent where $\psi(x)$ is the Polygamma function $\psi(x)= \frac{\Gamma'(u)}{\Gamma(u)}$. Near the singularity we have

\be
\frac{g^*_n (\tau)}{f_n (\tau)}\sim e^{i\zeta_n}e^{\pi n} \,,
\ee

\noindent where  $\zeta_n= \pi+2\gamma+2n\ln(f/2)$. For the fermionic part, we need to analyze $ \Pi e^{i\psi_n}\tan\varphi_n$. Using the  result (\ref{phi}) in the asymptotic limit, Equation (\ref{fdirac}), we get

\be
\Pi e^{i\psi_n}\tan\varphi_n\sim \Pi e^{i\psi'_n} e^{-\pi n} \,,
\ee

\noindent where $\psi'_n$ is the asymptotic limit of $\psi_n$. The entanglement state, in the asymptotic limit, takes the form

\begin{equation}
\lim_{\tau\rightarrow -\infty}|0(\tau),p_-\rangle= |0(\beta),p_-\rangle=\frac{1}{\sqrt{Z}}\prod_{n=1}^{\infty} e^{e^{-n\frac{\beta}{2}+i\zeta_n}\alpha^{\dagger}_n\cdot\tilde{\alpha}^\dagger_n
+e^{\Pi e^{-n\frac{\beta}{2}+i\psi'_n}}S^{\dagger}_n\cdot\tilde{S}^\dagger_n}|0\rangle \,,
\label{tstate}
\end{equation}

\noindent where 

\begin{equation}
Z= \prod_{n=1}^{\infty} \frac{1}{1-e^{-\beta n}}\frac{1}{e^{\beta n}-1} \,,
\end{equation}

\noindent and $\beta = 2\pi$. Note that the phases $\zeta_n$ and $\psi^{'}_n$, as well as the matrix $\Pi$, do not alter the expected values. In this limit, up to a phase, the entanglement state is exactly a thermal state for 2d bosons and fermions at equilibrium temperature $T=\frac{1}{2\pi}$ (in natural units) \cite{ume1, ume2}, \cite{ChuUme}. Indeed, the expected values of the bosonic and fermionic number operators in the state (\ref{tstate}) are

\begin{eqnarray}
\langle 0(\beta), p_- |a_n a^{\dagger}_n |0(\beta), p_- \rangle &=& \frac{1}{e^{\beta n}-1} \,,
 \nonumber \\
\langle 0(\beta),p_- | S_n S_n^{\dagger} |0(\beta), p_- \rangle &=& \frac{1}{e^{\beta n}+1} \,,
\end{eqnarray}

\noindent which are precisely the Bose-Einstein and Fermi-Dirac distributions. 

Regarding the bosonic two-point function, the correlation between the left and right mode observed in the vacuum $|0(\tau),p_- \rangle$ allows to express the two-point function in a matrix representation of the form

\begin{displaymath}
\mathbf{G}(\sigma,\sigma',\tau)=\left|\begin{array}{cc}
G_{++}(\sigma,\sigma',\tau)& G_{+-}(\sigma,\sigma',\tau)\\
G_{-+}(\sigma,\sigma',\tau)& G_{--}(\sigma,\sigma',\tau)\end{array}\right| \,,
\end{displaymath}

\noindent where

\be
G_{++}(\sigma,\sigma',\tau)&=&\left\langle 0(\tau),p_-|X_R(\sigma,\tau)X_R(\sigma',\tau)|0(\tau),p_-\right\rangle \,, \nonumber \\
G_{+-}(\sigma,\sigma',\tau)&=&\left\langle 0(\tau),p_-|X_R(\sigma,\tau)X_L(\sigma',\tau)|0(\tau),p_-\right\rangle \,, \nonumber \\
G_{-+}(\sigma,\sigma',\tau)&=&\left\langle 0(\tau),p_-|X_L(\sigma,\tau)X_R(\sigma',\tau)|0(\tau),p_-\right\rangle \,, \nonumber \\
G_{--}(\sigma,\sigma',\tau)&=&\left\langle 0(\tau),p_-|X_L(\sigma,\tau)X_L(\sigma',\tau)|0(\tau),p_-\right\rangle \,, 
\ee

\noindent Near to the singularity,

\begin{equation}
G_{++}(\sigma,\sigma',\tau)= G_0 +\sum_{n=1}^{\infty}\frac{\lim_{\tau\rightarrow-\infty}|Z(\tau)|^2}{n}\left[e^{-i(\sigma-\sigma')n}+\frac{\cos n(\sigma-\sigma')}{e^{\beta n}-1}\right] \,,
\end{equation}

\noindent where $G_0$ is the zero mode part ($G_0\sim \frac{\cos(u')\sin(u')}{u}$). One can see that the two-point function has the same form as the two-point function of a real time finite temperature quantum field theory, with one of the closed string sector (left or right) playing the role of degrees of freedom of the thermal bath \cite{lands}, \cite{ume1, ume2}.

Now, let's analyze the entropy's behavior. In order to do it as $\tau \rightarrow -\infty$ and $\tau \rightarrow \infty$, we need the asymptotic behavior of the Bessel functions. For the flat space limit ($\tau \rightarrow \infty$), equation (\ref{Oflat}) gives $|g_{n}(\tau)|^2 \sim 0$, $\sin^2\varphi_n\sim 0$ and the entropy is zero just as expected.   Close to the singularity ($\tau \rightarrow -\infty$), using equations (\ref{OBC}) and (\ref{OFC}), the left/right entanglement entropy is

\begin{eqnarray}
\lim_{\tau\rightarrow -\infty} {\cal S}(\tau) &=& S_{singularity}=-8\sum_{n=1}^{\infty} \left[\frac{1}{e^{\beta n}-1}\ln \left( \frac{1}{e^{\beta n}-1}\right)
-\frac{1}{1-e^{-\beta n}}\ln \left(\frac{1}{1-e^{-\beta n}}\right) \right] \nonumber \\
&-&8\sum_{n=1}^{\infty} \left[\frac{1}{e^{\beta n}+1}\ln \left( \frac{1}{e^{\beta n}+1}\right)
-\frac{1}{1+e^{-\beta n}}\ln \left(\frac{1}{1+e^{-\beta n}}\right)\right] \,,
\label{ssing}
\end{eqnarray}

\noindent Remark that, near the singularity, $S$ does not depend on time and it is not affected by the divergence of the curvature at $\tau\longrightarrow -\infty $. Actually, $S$ is finite at the singularity and it is exactly equal to the thermodynamic entropy of a 2d bosonic/fermionic gas at an equilibrium temperature $T = \frac{1}{2\pi}$.  Remember that the string coupling diverges in this limit and the relevant degrees of freedom belong to the non perturbative sector of the string theory. Moreover, the thermal state is the one of an open superstring; in fact, the left movers are traced out in the entropy calculation.

\section{Conclusion}

We have investigated new issues in the Ramond-Ramond time dependent superstring sigma model studied in \cite{Bin}. We have constructed a Bogoliubov generator that relates a time independent vacuum with a time dependent one ($\left|0(\tau),p_-\right\rangle$). This transformation can be interpreted as a relation between asymptotically flat observers and observers at a finite time; the late ones go with the string towards to the singularity. From the point of view of the asymptotically flat observers, the vacuum $\left|0(\tau),p_-\right\rangle$ is a superposition of $SU(1, 1)\times SU(2)$ coherent
states. Actually, it was shown that the Bogoliubov transformed vacuum is a time dependent left/right entanglement state.

 We have carried out the mode summation for the equal time superstring two-point function and presented the results in terms of Bessel, Hypergeometric and q-Polygamma functions. The behavior of the bosonic two-point function is different for each vacuum. In the time independent vacuum, it was investigated two regimes: close to flat space limit and close to singularity. In the first one, we have shown that the behavior at short distances of the two-point function is the same as that of the 
flat space. In the second it was shown that the short distance behavior of the two-point function close to background singularity is not singular; in particular, close to the singularity the bosonic two-point function goes to zero. This may corroborate the idea that the string gets highly excited and breaks up into bits propagating independently near the singularity, as it was argued in \cite{Madhu}. However, as in this background the string coupling gets higher at the singularity, we need to take into account non perturbative effects in order to have the exactly picture of the two-point function close to singularity. For the Bogoliubov transformed vacuum, in the bosonic sector we do not have the separation between the temporal and the ($\sigma-\sigma'$) dependence; we have two different behaviors separated by the similar point ${\tilde f}|\sigma-\sigma'| \sim e^{\tau}$.  For $\frac{{\tilde f}|\sigma-\sigma'|}{e^{\tau}}<<1$, the leading short-distance behavior of  the two-point function is again the same as the flat space. For $\frac{{\tilde f}|\sigma-\sigma'|}{e^{\tau}}>>1$ we have an exponential damping term. For the fermionic sector, in both vacuums the two-point function is written in terms of q-Polygamma functions and the behavior at short distances is not altered by the background.

The structure of the two-point function in the Bogoliubov transformed vaccum is very similar to the one of a thermal theory, with one of the closed string sector (left or right) playing the role of degrees of freedom of the thermal bath; this fact is corroborated when analyzing the left/right entanglement entropy. Although the Hamiltonian diverges as the string approaches the cosmological singularity, the left/right entanglement entropy is well-behaved and becomes a thermodynamic entropy. Actually, it was shown that, close to the singularity, the finite time vacuum is unitarily inequivalent to the asymptotically flat vacuum. In reference \cite{dafdani08} (where the same problem is analized for the model studied in \cite{PRT}), this non unitarity is related to the fact that, close to the cosmological singularity, for asymptotically flat observers, the closed string vacuum appears as a D-brane described in the closed string channel. Here, the closed superstring vacuum appears as an open superstring thermal vacuum. In particular, it was shown that the non unitary dynamics is governed by an entropy operator, similar to what happens in quantum dissipation theory  \cite{ceravi}. As a future work, it will be interesting to study this thermalization from the perspective of the Eigenstate Thermalization Hypothesis in Conformal Field Theory discussed in \cite{Lash}, as well as to investigate if this kind of thermalization appears in time dependent orbifold models. Finally, it was presented in the appendix how to take into account a non equilibrium thermalization for the zero mode of the GS superstring. It will be interesting to extend this analysis for the non zero string modes and have a toy model to study the non equilibrium thermodynamics of the superstring close to null singularities.

\appendix

\section{Zero mode thermalization}

As the interaction of the string modes with the background provides a thermal bath for the string, we need to deal with the thermalization of the zero mode. It was shown that thermalization occurs close to the singularity, which implies an immediate problem: the time dependent frequency diverges close to the singularity. So, in order to take care of the zero mode thermalization, we need to impose a cut-off on the frequency. As the string coupling diverges close to the singularity, we can speculate that some non perturbative effect of string theory may account for the divergence in frequency. Clearly, this is a kind of  "God ex machina" solution and this appendix is somewhat speculative. In the meantime, we show how to deal with the zero mode in non equilibrium thermodynamics. 

Let us start with the bosonic sector. The zero mode has the following time dependent Hamiltonian

\begin{equation}
{\cal H}= p^+H_{0}(\tau)=
    \frac{1}{2}\left[(p^I_0)^2+\tilde{f}^2e^{-2\tau}\left(\frac{x^I_0}{\alpha'}\right)^2\right] \,,
\end{equation}

\noindent with

\begin{equation}
[\tilde{x}^I,\,\tilde{p}^J]=i\,\delta^{IJ} \,.
\label{xp}
\end{equation}

\noindent We are going to use ${\cal H}$ as time evolution operator.  In order to take care of  the zero mode thermalization, one could naively construct a thermal density matrix, defined by the time dependent Hamiltonian

\begin{equation}
\rho_H=\frac{1}{Z}e^{-\beta H} \,.
\label{rhoh}
\end{equation}

\noindent This density matrix does not satisfy the quantum Liouville-von Neumann (LvN) equation and it is difficult to relate $1/\beta$ to the equilibrium temperature. If the system starts in the initial thermal equilibrium state, its final state can be far away from the initial one; one way to solve this problem is given by the LvN approach. The essential idea of the LvN method is that the quantum LvN equation provides all the quantum and statistical information of non equilibrium systems. The strategy of this approach is to define time dependent oscillators that satisfy the equation \cite{Kim,KMMS,Lewis}

\begin{equation}
i\frac{\partial a}{\partial \tau} +[a,H] =0 \,.
\label{LN}
\end{equation} 

\noindent The linearity of the LvN equation allows us to use ${a}(\tau)$ and
${a}^{\dagger}(\tau)$ to construct operators that also satisfy Equation (\ref{LN}); in particular, the number and the density operator. By defining the number operator in the usual way

\begin{equation}
\hat{N} (\tau) = {a}^{\dagger} (\tau) {a} (\tau) \,,
\end{equation}

\noindent one finds the Fock space consisting of the time dependent number
states

\begin{equation}
\hat{N} (\tau) \vert n, \tau \rangle = n \vert n, \tau \rangle \,.
\label{numst}
\end{equation}

\noindent The time dependent zero mode vacuum state is the one that is annihilated by ${a} (\tau)$ and the $n$-th number state is obtained by applying ${a}^{\dagger}(\tau)$ $n$-times:

\begin{eqnarray}
{a} (\tau) \vert 0, \tau \rangle = 0 \,, \nonumber\\ 
\vert n, \tau \rangle
= \frac{({a}^{\dagger} (\tau))^n}{\sqrt{n!}} \vert 0, \tau \rangle \,.
\end{eqnarray}

\noindent With these oscillators, a density matrix of the thermal type (which satisfies the LvN equation) can be defined as

\begin{equation}
\rho_{\rm T} = e^{\beta\omega_0 a^{\dagger}(\tau)a(\tau)} \,,
\label{rhoT}
\end{equation}

\noindent where $\beta$ and $\omega_0$ are free parameters and $Z_N$ is the partition function given by

\begin{equation}
Z_N = \sum_{n = 0}^{\infty} \langle n, \tau \vert e^{- \beta 
\omega_0 ({N} (\tau) + \frac{1}{2})} \vert n, \tau \rangle =
\frac{1}{2\sinh(\frac{\beta  \omega_0}{2})} \,.
\end{equation}

\noindent Equation (\ref{rhoT}) has the same form as the standard density operator, with the time independent annihilation and creation operators being replaced by the time dependent ones. So Equation (\ref{rhoT}) includes the time independent case as a special case when one chooses $\beta = 1/( T)$ and $\omega_0$ as the oscillator frequency at the equilibrium temperature. For our case, $\beta =2\pi$ is fixed by the asymptotic behavior of the entanglement state and $\omega_0$ will be the cut-off frequency. In this way, the LvN method treats the time dependent, non equilibrium system exactly in the same way as the time independent, equilibrium one.
 
Let's apply this procedure to the bosonic zero mode by defining the following time dependent creation/annihilation operators (from now on, we won't use the spacetime indices)

\begin{eqnarray}
a(\tau)&=& \left(\phi^*(\tau)p - \dot\phi^*(\tau) x\right) \,, \nonumber \\
a^{\dagger}(\tau)&=& \left(\phi(\tau)p - \dot\phi(\tau) x\right)  \,.
\label{osci}
\end{eqnarray}

\noindent The commutation relation 

\begin{equation}
[a,a^{\dagger}]=1 \,,
\end{equation}

\noindent is obtained from (\ref{xp}) if $\phi (\tau)$ satisfies the Wronskian

\begin{equation}
\dot{\phi}^*\phi - \phi^*\dot{\phi}= i \,.
\label{Wr}
\end{equation}

\noindent By placing (\ref{osci}) in (\ref{LN}), we obtain

\begin{eqnarray}
\frac{d^2\phi}{d\tau^2}&+& \omega^2(\tau) \phi(\tau)=0 \,,
\label{eqm}
\end{eqnarray}

\noindent where $\omega(\tau) = \tilde{f}e^{-\tau}$. The solution for $\phi (\tau)$ that satisfies the Equations (\ref{eqm}) and (\ref{Wr}) has the form

\begin{equation}
\phi(\tau) = \frac{\sqrt{\pi}}{2}\left[J_0(z) +iY_0(z)\right] \,,
\end{equation}

\noindent where $z = \omega (\tau)$ and the following property of Bessel functions was used

\begin{equation}
J(z)\dot{Y}(z)-\dot{J}(z)Y(z)= \frac{2}{\pi z} \,.
\end{equation}

\noindent In terms of the operators defined in (\ref{osci}), the momentum and position are written as

\begin{eqnarray}
p&=& i\left[\dot{\phi}(\tau)a-\dot{\phi}^*(\tau)a^{\dagger}\right] \,, \nonumber \\
x&=& i\left[{\phi}(\tau)a-{\phi}^*(\tau)a^{\dagger}\right] \,,
\end{eqnarray}

\noindent and the Hamiltonian is 

\begin{equation}
H= \frac{1}{2p^+}\left[2\Omega_0\left(a^{\dagger}(\tau)a(\tau)+ d\right) - C_0 a^2(\tau) -C^*_0 (a^\dagger)^2(\tau)\right] \,,
\end{equation}

\noindent where

\begin{eqnarray}
\Omega_0= |\dot{\phi}(\tau)|^2+\omega^2(\tau)|\phi(\tau)|^2 \,, \nonumber \\
C_0 = {\dot{\phi}}^2+ \omega^2(\tau){\phi(\tau)}^2 \,.
\end{eqnarray}

\noindent Perceive the similarity with the Hamiltonian for non zero modes.

\subsection{Density matrix}
 
 We will now construct the coordinate representation of the density matrix in terms of the invariant oscillators (\ref{osci}) and compare it with the density matrix defined in (\ref{rhoh}). In the coordinate representation the number state is given by 

\begin{equation}
\Psi_n (x, \tau) = \Biggl(\frac{1}{2 \pi \phi^* (\tau) \phi (\tau)}
\Biggr)^{1/4} \frac{1}{\sqrt{2^n n!}} \Biggl(\frac{\phi(\tau)}{\phi^* (\tau)}
\Biggr)^n H_n \Biggl(\frac{x}{\sqrt{2\phi^*(\tau) \phi (\tau)}}
\Biggr) \exp \Biggl[\frac{i}{2}\frac{m}{\hbar}
\frac{\dot{\phi}^*(\tau)}{\phi^*(\tau)} x^2 \Biggr] \,, 
\label{har wav}
\end{equation}

\noindent where the $H_n$ are the Hermite polynomials. The matrix elements of the density matrix (\ref{rhoT}) in coordinate representation are given by 

\begin{eqnarray}
\rho_{\rm T} (x', x, \tau) &=& \left\langle x|\rho|x'\right\rangle=\frac{1}{Z_N} \sum_{n = 0}^{\infty}
\Psi_n (x', \tau) \Psi^*_n (x, \tau) e^{- \beta  \omega_0 (n +
\frac{1}{2})}\nonumber\\ 
&=& A(x,x',\tau) \Biggl[\frac{\tanh(\frac{\beta 
\omega_0}{2})}{2 \pi  \phi^* \phi}\Biggr]^{1/2} \nonumber\\ 
&\times& \exp \Biggl[-\frac{1}{8 \phi^*
\phi} \Biggl\{ (x' + x)^2 \tanh\left(\frac{\beta \omega_0}{2}\right) +
(x'-x)^2 \coth\left(\frac{\beta \omega_0}{2}\right) \Biggr\} \Biggr] \,, \nonumber\\
\label{cden}
\end{eqnarray}

\noindent where we have used

\begin{equation}
H_n (x') H_n (x) = \frac{1}{\pi} e^{x'^2 + x^2} \int_{-
\infty}^{\infty} \int_{- \infty}^{\infty} dz_1 dz_2 (2iz_1)^n (2 i
z_2)^n e^{- z_1^2 - 2 i x' z_1 - z_2^2 - 2 i x z_2} \,,
\end{equation}

\noindent and $A(x,x',\tau)$ is 

\begin{equation}
A(x,x',\tau)=e^{\frac{i}{2}\frac{\dot{\phi}^*}{\phi^*}x'^2-\frac{i}{2}\frac{\dot{\phi}}{\phi}x^2}e^{\frac{\omega(\tau)}{2}\left(x'^2+x^2\right)} \,.
\end{equation}

On the other hand, the density operator (\ref{rhoh}) for the instantaneous Hamiltonian has the matrix representation

\begin{eqnarray}
\rho_{\rm H} (x', x, \tau) &=& \Biggl[\frac{\omega (\tau)
\tanh(\frac{\beta  \omega (\tau)}{2})}{\pi }\Biggr]^{1/2}
\nonumber\\ &\times& \exp \Biggl[-\frac{\omega (\tau)}{4}
\Biggl\{ (x' + x)^2 \tanh\left(\frac{\beta \omega (\tau)}{2}\right) +
(x'-x)^2 \coth\left(\frac{\beta \omega (\tau)}{2}\right) \Biggr\} \Biggr] \,.\nonumber\\
\label{den mat3}
\end{eqnarray}

\noindent At the asymptotic limit ($\tau\rightarrow -\infty$) where thermalization occurs, we have

\begin{eqnarray}
\phi\phi^* &\rightarrow& \frac{1}{2\tilde{f}e^{-\tau}}=\frac{1}{2\omega(\tau)} \,,\nonumber \\
\frac{\dot{\phi}}{\phi} &\rightarrow&  -i\omega(\tau) \,,\nonumber \\
A(x,x',\tau)&\rightarrow&  1 \,.
\end{eqnarray}

\noindent The density matrix (\ref{cden}) becomes 

\begin{eqnarray}
\rho_{\rm T} (x', x, \tau) &=& \Biggl[\omega(\tau)\tanh\left(\frac{\beta 
\omega_0}{2}\right)\Biggr]^{1/2} \nonumber\\ 
&\times& \exp \Biggl[-\frac{\omega(\tau)}{4} \Biggl\{ (x' + x)^2 \tanh\left(\frac{\beta \omega_0}{2}\right) +
(x'-x)^2 \coth\left(\frac{\beta \omega_0}{2}\right) \Biggr\} \Biggr] \,. \nonumber\\
\end{eqnarray}

\noindent So, as $\omega(\tau)$ gets close to $\omega_0$, $\rho_{\rm T}$ gets close to $\rho_{H}$.

\subsection{Fermionic Sector}

The zero mode of the fermionic Hamiltonian has the form

\begin{equation}
{\cal H}_f = p^+ H_{F0} =  -\frac{2\,i}{\alpha'} \tilde{f} e^{-\tau}S_0 \Pi \tilde{S}_0 \,.
\end{equation}

\noindent Let us define (without using spacetime indices)

\begin{eqnarray}
S_R&=& \frac{1}{\sqrt{2}}\left(1+\Pi\right)\left(S_0+i\tilde{S}_0\right),\:\: S_R^{\dagger}= \frac{1}{\sqrt{2}}\left(1+\Pi\right)\left(S_0-i\tilde{S}_0\right) \,, \nonumber \\
S_L&=&\frac{1}{\sqrt{2}}\left(1-\Pi\right)\left(S_0+i\tilde{S}_0\right),\:\: S_L^{\dagger}= \frac{1}{\sqrt{2}}\left(1-\Pi\right)\left(S_0-i\tilde{S}_0\right) \,,
\end{eqnarray}

\noindent where the fermionic  operators $ S_L $ and $ S_R $ have 4 components each  and satisfy

\begin{eqnarray}
\{S_R,S_R^{\dagger}\}&=& \{S_L,S_L^{\dagger}\}=1 \,, \nonumber \\
\{S_R,S_R\}&=& \{S_R^{\dagger},S_R^{\dagger}\}=0 \,, \nonumber \\
\{S_L,S_L\}&=& \{S_L^{\dagger},S_L^{\dagger}\}=0 \,.
 \end{eqnarray}

\noindent The Hamiltonian takes the form

\begin{equation}
H_{F0}= \omega(\tau)\left( S_R^{\dagger}S_R- S_L^{\dagger}S_L -8\right) \,,
\end{equation}

\noindent where $ \omega (\tau) = {\tilde f}e^{-\tau} $. The term $ - 8 $ comes from the normal ordering and cancels with the bosonic one. Following the LvN approach, we will define $ S_R (\tau) $ and $ S_L (\tau) $ such that they satisfy

\begin{eqnarray}
i\frac{\partial S_R(\tau)}{\partial \tau}+\left[S_R(\tau),H\right]&=&0 \,, \nonumber \\ 
i\frac{\partial S_L(\tau)}{\partial \tau}+\left[S_L(\tau),H\right]&=&0 \,,\nonumber \\
\{S_R(\tau),S_R^{\dagger}(\tau) \}&=& \{S_L(\tau),S_L^{\dagger}(\tau) \}=1\,.
\label{LNVf}
\end{eqnarray}
 
 \noindent Let us propose
 
\begin{eqnarray}
S_R(\tau)= F_+(\tau)S_R+G_+(\tau)S_L^{\dagger} \,, \nonumber \\
S_L(\tau)= F_-(\tau)S_L+G_-(\tau)S_R^{\dagger} \,.
\end{eqnarray}

\noindent The anti-commutation relations are guaranteed if

\begin{equation}
|F|^2+|G|^2=1 \,.
\label{Wf}
\end{equation}

\noindent The solutions of (\ref{LNVf}) that satisfy (\ref{Wf}) are

\begin{eqnarray}
F_{+}(\tau) = G_+(\tau) = \frac{1}{\sqrt{2}}e^{-i\omega(\tau)} \,, \nonumber \\
F_{-}(\tau) = G_-(\tau) = \frac{1}{\sqrt{2}}e^{i\omega(\tau)} \,.
\end{eqnarray}

\noindent In terms of $S_R(\tau)$ and $S_L(\tau) $ the Hamiltonian is

\begin{equation}
H= \omega(\tau)\left[ S_R(\tau)^\dagger S_R(\tau)  -S_L^{\dagger}(\tau)S_L(\tau) + S_R(\tau)^\dagger (\tau)S_L^{\dagger}(\tau)+S_L(\tau)S_R(\tau) \right] \,.
\end{equation}

\noindent The LvN invariant fermionic density matrix  is built in a similar way to the bosonic one. 

\bibliographystyle{JHEP}
\bibliography{biblio}

\providecommand{\href}[2]{#2}\begingroup\raggedright\begin{thebibliography}{10}

\bibitem{orb1}
A.~Arduino, R.~Finotello and I.~Pesando, \emph{{On the Origin of Divergences in
  Time-Dependent Orbifolds}},
  \href{https://arxiv.org/abs/2002.11306}{{\ttfamily 2002.11306}}.

\bibitem{orb2}
P.~Malkiewicz and W.~Piechocki, \emph{{Excited states of a string in time
  dependent orbifold}},
  \href{https://doi.org/10.1088/0264-9381/26/1/015008}{\emph{Class. Quant.
  Grav.} {\bfseries 26} (2009) 015008}
  [\href{https://arxiv.org/abs/0807.2990}{{\ttfamily 0807.2990}}].

\bibitem{orb3}
H.~Liu, G.~W. Moore and N.~Seiberg, \emph{{Strings in a time dependent
  orbifold}}, \href{https://doi.org/10.1088/1126-6708/2002/06/045}{\emph{JHEP}
  {\bfseries 06} (2002) 045}
  [\href{https://arxiv.org/abs/hep-th/0204168}{{\ttfamily hep-th/0204168}}].

\bibitem{orb4}
L.~Cornalba and M.~S. Costa, \emph{{A New cosmological scenario in string
  theory}}, \href{https://doi.org/10.1103/PhysRevD.66.066001}{\emph{Phys. Rev.
  D} {\bfseries 66} (2002) 066001}
  [\href{https://arxiv.org/abs/hep-th/0203031}{{\ttfamily hep-th/0203031}}].

\bibitem{orb5}
G.~T. Horowitz and J.~Polchinski, \emph{{Instability of space - like and null
  orbifold singularities}},
  \href{https://doi.org/10.1103/PhysRevD.66.103512}{\emph{Phys. Rev. D}
  {\bfseries 66} (2002) 103512}
  [\href{https://arxiv.org/abs/hep-th/0206228}{{\ttfamily hep-th/0206228}}].

\bibitem{orb6}
A.~Lawrence, \emph{{On the Instability of 3-D null singularities}},
  \href{https://doi.org/10.1088/1126-6708/2002/11/019}{\emph{JHEP} {\bfseries
  11} (2002) 019} [\href{https://arxiv.org/abs/hep-th/0205288}{{\ttfamily
  hep-th/0205288}}].

\bibitem{orb7}
H.~Liu, G.~W. Moore and N.~Seiberg, \emph{{Strings in time dependent
  orbifolds}}, \href{https://doi.org/10.1088/1126-6708/2002/10/031}{\emph{JHEP}
  {\bfseries 10} (2002) 031}
  [\href{https://arxiv.org/abs/hep-th/0206182}{{\ttfamily hep-th/0206182}}].

\bibitem{orb8}
M.~Berkooz, B.~Craps, D.~Kutasov and G.~Rajesh, \emph{{Comments on cosmological
  singularities in string theory}},
  \href{https://doi.org/10.1088/1126-6708/2003/03/031}{\emph{JHEP} {\bfseries
  03} (2003) 031} [\href{https://arxiv.org/abs/hep-th/0212215}{{\ttfamily
  hep-th/0212215}}].

\bibitem{orb9}
M.~Berkooz, Z.~Komargodski, D.~Reichmann and V.~Shpitalnik, \emph{{Flow of
  geometries and instantons on the null orbifold}},
  \href{https://doi.org/10.1088/1126-6708/2005/12/018}{\emph{JHEP} {\bfseries
  12} (2005) 018} [\href{https://arxiv.org/abs/hep-th/0507067}{{\ttfamily
  hep-th/0507067}}].

\bibitem{matr1}
B.~Craps and O.~Evnin, \emph{{Light-like Big Bang singularities in string and
  matrix theories}},
  \href{https://doi.org/10.1088/0264-9381/28/20/204006}{\emph{Class. Quant.
  Grav.} {\bfseries 28} (2011) 204006}
  [\href{https://arxiv.org/abs/1103.5911}{{\ttfamily 1103.5911}}].

\bibitem{matr2}
B.~Craps, S.~Sethi and E.~P. Verlinde, \emph{{A Matrix big bang}},
  \href{https://doi.org/10.1088/1126-6708/2005/10/005}{\emph{JHEP} {\bfseries
  10} (2005) 005} [\href{https://arxiv.org/abs/hep-th/0506180}{{\ttfamily
  hep-th/0506180}}].

\bibitem{matr3}
M.~Li, \emph{{A Class of cosmological matrix models}},
  \href{https://doi.org/10.1016/j.physletb.2005.08.099}{\emph{Phys. Lett. B}
  {\bfseries 626} (2005) 202}
  [\href{https://arxiv.org/abs/hep-th/0506260}{{\ttfamily hep-th/0506260}}].

\bibitem{matr4}
M.~Li and W.~Song, \emph{{Shock waves and cosmological matrix models}},
  \href{https://doi.org/10.1088/1126-6708/2005/10/073}{\emph{JHEP} {\bfseries
  10} (2005) 073} [\href{https://arxiv.org/abs/hep-th/0507185}{{\ttfamily
  hep-th/0507185}}].

\bibitem{tac1}
A.~Kostouki, \emph{{Tachyon-Dilaton driven Inflation as an alpha-prime -
  non-perturbative solution in First Quantized String Cosmology}},
  \href{https://doi.org/10.1088/1742-6596/171/1/012030}{\emph{J. Phys. Conf.
  Ser.} {\bfseries 171} (2009) 012030}
  [\href{https://arxiv.org/abs/0905.2552}{{\ttfamily 0905.2552}}].

\bibitem{tac2}
A.~Sen, \emph{{Rolling tachyon}},
  \href{https://doi.org/10.1088/1126-6708/2002/04/048}{\emph{JHEP} {\bfseries
  04} (2002) 048} [\href{https://arxiv.org/abs/hep-th/0203211}{{\ttfamily
  hep-th/0203211}}].

\bibitem{tac3}
A.~Sen, \emph{{Open closed duality at tree level}},
  \href{https://doi.org/10.1103/PhysRevLett.91.181601}{\emph{Phys. Rev. Lett.}
  {\bfseries 91} (2003) 181601}
  [\href{https://arxiv.org/abs/hep-th/0306137}{{\ttfamily hep-th/0306137}}].

\bibitem{tac4}
J.~McGreevy and E.~Silverstein, \emph{{The Tachyon at the end of the
  universe}}, \href{https://doi.org/10.1088/1126-6708/2005/08/090}{\emph{JHEP}
  {\bfseries 08} (2005) 090}
  [\href{https://arxiv.org/abs/hep-th/0506130}{{\ttfamily hep-th/0506130}}].

\bibitem{berkoos}
M.~Berkooz and D.~Reichmann, \emph{{A Short Review of Time Dependent Solutions
  and Space-like Singularities in String Theory}},
  \href{https://doi.org/10.1016/j.nuclphysbps.2007.06.008}{\emph{Nucl. Phys. B
  Proc. Suppl.} {\bfseries 171} (2007) 69}
  [\href{https://arxiv.org/abs/0705.2146}{{\ttfamily 0705.2146}}].

\bibitem{pbb}
G.~Veneziano, \emph{{String cosmology: The Pre - big bang scenario}},  in
  \emph{{The primordial universe. Proceedings, Summer School on physics, 71st
  session, Les Houches, France, June 28-July 23, 1999}}, pp.~581--628, 2000,
  \href{https://doi.org/10.1007/3-540-45334-2\_12}{DOI}
  [\href{https://arxiv.org/abs/hep-th/0002094}{{\ttfamily hep-th/0002094}}].

\bibitem{HS1}
G.~T. Horowitz and A.~R. Steif, \emph{{Strings in Strong Gravitational
  Fields}}, \href{https://doi.org/10.1103/PhysRevD.42.1950}{\emph{Phys. Rev. D}
  {\bfseries 42} (1990) 1950}.

\bibitem{HS2}
G.~T. Horowitz and A.~R. Steif, \emph{{Space-Time Singularities in String
  Theory}}, \href{https://doi.org/10.1103/PhysRevLett.64.260}{\emph{Phys. Rev.
  Lett.} {\bfseries 64} (1990) 260}.

\bibitem{RB}
R.~Brooks, \emph{{Plane wave gravitons, curvature singularities and string
  physics}}, \href{https://doi.org/10.1142/S0217732391000877}{\emph{Mod. Phys.
  Lett. A} {\bfseries 6} (1991) 841}.

\bibitem{VS1}
H.~de~Vega and N.~G. Sanchez, \emph{{Strings falling into space-time
  singularities}}, \href{https://doi.org/10.1103/PhysRevD.45.2783}{\emph{Phys.
  Rev. D} {\bfseries 45} (1992) 2783}.

\bibitem{VS2}
H.~de~Vega, M.~Ramon~Medrano and N.~G. Sanchez, \emph{{Classical and quantum
  strings near space-time singularities: Gravitational plane waves with
  arbitrary polarization}},
  \href{https://doi.org/10.1088/0264-9381/10/10/008}{\emph{Class. Quant. Grav.}
  {\bfseries 10} (1993) 2007}.

\bibitem{VS3}
N.~G. Sanchez, \emph{{Classical and quantum strings in plane waves, shock waves
  and space-time singularities: Synthesis and new results}},
  \href{https://doi.org/10.1142/S0217751X03015787}{\emph{Int. J. Mod. Phys. A}
  {\bfseries 18} (2003) 4797}
  [\href{https://arxiv.org/abs/hep-th/0302214}{{\ttfamily hep-th/0302214}}].

\bibitem{BFHP1}
M.~Blau, J.~M. Figueroa-O'Farrill, C.~Hull and G.~Papadopoulos, \emph{{A New
  maximally supersymmetric background of IIB superstring theory}},
  \href{https://doi.org/10.1088/1126-6708/2002/01/047}{\emph{JHEP} {\bfseries
  01} (2002) 047} [\href{https://arxiv.org/abs/hep-th/0110242}{{\ttfamily
  hep-th/0110242}}].

\bibitem{BFHP2}
M.~Blau, J.~M. Figueroa-O'Farrill, C.~Hull and G.~Papadopoulos, \emph{{Penrose
  limits and maximal supersymmetry}},
  \href{https://doi.org/10.1088/0264-9381/19/10/101}{\emph{Class. Quant. Grav.}
  {\bfseries 19} (2002) L87}
  [\href{https://arxiv.org/abs/hep-th/0201081}{{\ttfamily hep-th/0201081}}].

\bibitem{BFP}
M.~Blau, J.~M. Figueroa-O'Farrill and G.~Papadopoulos, \emph{{Penrose limits,
  supergravity and brane dynamics}},
  \href{https://doi.org/10.1088/0264-9381/19/18/310}{\emph{Class. Quant. Grav.}
  {\bfseries 19} (2002) 4753}
  [\href{https://arxiv.org/abs/hep-th/0202111}{{\ttfamily hep-th/0202111}}].

\bibitem{RP}
R.~Penrose, \emph{Any space-time has a plane wave as a limit},  in
  \emph{Differential Geometry and Relativity}, Springer, Dordrecht, (1976).

\bibitem{RG}
R.~Gueven, \emph{{Plane wave limits and T duality}},
  \href{https://doi.org/10.1016/S0370-2693(00)00517-7}{\emph{Phys. Lett. B}
  {\bfseries 482} (2000) 255}
  [\href{https://arxiv.org/abs/hep-th/0005061}{{\ttfamily hep-th/0005061}}].

\bibitem{RM}
R.~Metsaev, \emph{{Type IIB Green-Schwarz superstring in plane wave
  Ramond-Ramond background}},
  \href{https://doi.org/10.1016/S0550-3213(02)00003-2}{\emph{Nucl. Phys. B}
  {\bfseries 625} (2002) 70}
  [\href{https://arxiv.org/abs/hep-th/0112044}{{\ttfamily hep-th/0112044}}].

\bibitem{MT}
R.~Metsaev and A.~A. Tseytlin, \emph{{Exactly solvable model of superstring in
  Ramond-Ramond plane wave background}},
  \href{https://doi.org/10.1103/PhysRevD.65.126004}{\emph{Phys. Rev. D}
  {\bfseries 65} (2002) 126004}
  [\href{https://arxiv.org/abs/hep-th/0202109}{{\ttfamily hep-th/0202109}}].

\bibitem{BMN}
D.~E. Berenstein, J.~M. Maldacena and H.~S. Nastase, \emph{{Strings in flat
  space and pp waves from N=4 superYang-Mills}},
  \href{https://doi.org/10.1088/1126-6708/2002/04/013}{\emph{JHEP} {\bfseries
  04} (2002) 013} [\href{https://arxiv.org/abs/hep-th/0202021}{{\ttfamily
  hep-th/0202021}}].

\bibitem{PRT}
G.~Papadopoulos, J.~Russo and A.~A. Tseytlin, \emph{{Solvable model of strings
  in a time dependent plane wave background}},
  \href{https://doi.org/10.1088/0264-9381/20/5/313}{\emph{Class. Quant. Grav.}
  {\bfseries 20} (2003) 969}
  [\href{https://arxiv.org/abs/hep-th/0211289}{{\ttfamily hep-th/0211289}}].

\bibitem{Bin}
B.~Chen, Y.-l. He and P.~Zhang, \emph{Exactly solvable model of superstring in
  plane-wave background with linear null dilaton},
  \href{https://doi.org/10.1016/j.nuclphysb.2006.02.019}{\emph{Nucl. Phys.}
  {\bfseries B741} (2006) 269}
  [\href{https://arxiv.org/abs/hep-th/0509113}{{\ttfamily hep-th/0509113}}].

\bibitem{DipDas}
D.~Das and S.~Datta, \emph{{Universal features of left-right entanglement
  entropy}}, \href{https://doi.org/10.1103/PhysRevLett.115.131602}{\emph{Phys.
  Rev. Lett.} {\bfseries 115} (2015) 131602}
  [\href{https://arxiv.org/abs/1504.02475}{{\ttfamily 1504.02475}}].

\bibitem{Ryang}
S.-j. Ryang, \emph{{String propagators in time dependent and time independent
  homogeneous plane waves}},
  \href{https://doi.org/10.1088/1126-6708/2003/11/007}{\emph{JHEP} {\bfseries
  11} (2003) 007} [\href{https://arxiv.org/abs/hep-th/0310044}{{\ttfamily
  hep-th/0310044}}].

\bibitem{Balasu}
V.~Balasubramanian, M.~B. McDermott and M.~Van~Raamsdonk, \emph{{Momentum-space
  entanglement and renormalization in quantum field theory}},
  \href{https://doi.org/10.1103/PhysRevD.86.045014}{\emph{Phys. Rev. D}
  {\bfseries 86} (2012) 045014}
  [\href{https://arxiv.org/abs/1108.3568}{{\ttfamily 1108.3568}}].

\bibitem{GMN}
A.~Gadelha, D.~Z. Marchioro and D.~L. Nedel, \emph{{Entanglement and entropy
  operator for strings in pp-wave time dependent background}},
  \href{https://doi.org/10.1016/j.physletb.2006.06.044}{\emph{Phys. Lett. B}
  {\bfseries 639} (2006) 383}
  [\href{https://arxiv.org/abs/hep-th/0605237}{{\ttfamily hep-th/0605237}}].

\bibitem{z1}
L.~A. Pando~Zayas and N.~Quiroz, \emph{{Left-Right Entanglement Entropy of
  Boundary States}}, \href{https://doi.org/10.1007/JHEP01(2015)110}{\emph{JHEP}
  {\bfseries 01} (2015) 110} [\href{https://arxiv.org/abs/1407.7057}{{\ttfamily
  1407.7057}}].

\bibitem{z2}
L.~A. Pando~Zayas and N.~Quiroz, \emph{{Left-Right Entanglement Entropy of
  Dp-branes}}, \href{https://doi.org/10.1007/JHEP11(2016)023}{\emph{JHEP}
  {\bfseries 11} (2016) 023}
  [\href{https://arxiv.org/abs/1605.08666}{{\ttfamily 1605.08666}}].

\bibitem{Calabre}
P.~Calabrese and J.~L. Cardy, \emph{{Time-dependence of correlation functions
  following a quantum quench}},
  \href{https://doi.org/10.1103/PhysRevLett.96.136801}{\emph{Phys. Rev. Lett.}
  {\bfseries 96} (2006) 136801}
  [\href{https://arxiv.org/abs/cond-mat/0601225}{{\ttfamily
  cond-mat/0601225}}].

\bibitem{Rigol}
M.~Rigol, V.~Dunjko, V.~Yurovsky and M.~Olshanii, \emph{{Relaxation in a
  Completely Integrable Many-Body Quantum System: An Ab Initio Study of the
  Dynamics of the Highly Excited States of 1D Lattice Hard-Core Bosons}},
  \href{https://doi.org/10.1103/PhysRevLett.98.050405}{\emph{Phys. Rev. Lett.}
  {\bfseries 98} (2007) 050405}
  [\href{https://arxiv.org/abs/cond-mat/0604476}{{\ttfamily
  cond-mat/0604476}}].

\bibitem{Kim}
S.~P. Kim and C.~H. Lee, \emph{{Nonequilibrium quantum dynamics of second order
  phase transitions}},
  \href{https://doi.org/10.1103/PhysRevD.62.125020}{\emph{Phys. Rev.}
  {\bfseries D62} (2000) 125020}
  [\href{https://arxiv.org/abs/hep-ph/0005224}{{\ttfamily hep-ph/0005224}}].

\bibitem{KMMS}
F.~C. Khanna, A.~P. Malbouisson, J.~M. Malbouisson and A.~R. Santana,
  \emph{{Thermal quantum field theory - Algebraic aspects and applications}}.
  World Scientific Publishing Company, 2009.

\bibitem{Lewis}
H.~R. Lewis and W.~B. Riesenfeld, \emph{{An Exact quantum theory of the time
  dependent harmonic oscillator and of a charged particle time dependent
  electromagnetic field}}, \href{https://doi.org/10.1063/1.1664991}{\emph{J.
  Math. Phys.} {\bfseries 10} (1969) 1458}.

\bibitem{Metsaev}
R.~Metsaev, \emph{{Type IIB Green-Schwarz superstring in plane wave
  Ramond-Ramond background}},
  \href{https://doi.org/10.1016/S0550-3213(02)00003-2}{\emph{Nucl. Phys. B}
  {\bfseries 625} (2002) 70}
  [\href{https://arxiv.org/abs/hep-th/0112044}{{\ttfamily hep-th/0112044}}].

\bibitem{Sadri}
D.~Sadri and M.~M. Sheikh-Jabbari, \emph{{The Plane wave / superYang-Mills
  duality}}, \href{https://doi.org/10.1103/RevModPhys.76.853}{\emph{Rev. Mod.
  Phys.} {\bfseries 76} (2004) 853}
  [\href{https://arxiv.org/abs/hep-th/0310119}{{\ttfamily hep-th/0310119}}].

\bibitem{gadcris}
M.~C.~B. Abdalla and A.~L. Gadelha, \emph{{General unitary SU(1,1) TFD
  formulation}},
  \href{https://doi.org/10.1016/j.physleta.2003.12.025}{\emph{Phys. Lett.}
  {\bfseries A322} (2004) 31}
  [\href{https://arxiv.org/abs/hep-th/0309254}{{\ttfamily hep-th/0309254}}].

\bibitem{gad1}
M.~C.~B. Abdalla, A.~L. Gadelha and I.~V. Vancea, \emph{{On the SU(1,1) thermal
  group of bosonic strings and D-branes}},
  \href{https://doi.org/10.1103/PhysRevD.66.065005}{\emph{Phys. Rev.}
  {\bfseries D66} (2002) 065005}
  [\href{https://arxiv.org/abs/hep-th/0203222}{{\ttfamily hep-th/0203222}}].

\bibitem{gad2}
M.~C.~B. Abdalla, A.~L. Gadelha and D.~L. Nedel, \emph{{On the entropy operator
  for the general SU(1,1) TFD formulation}},
  \href{https://doi.org/10.1016/j.physleta.2004.11.025}{\emph{Phys. Lett.}
  {\bfseries A334} (2005) 123}
  [\href{https://arxiv.org/abs/hep-th/0409116}{{\ttfamily hep-th/0409116}}].

\bibitem{gad3}
M.~C.~B. Abdalla, A.~L. Gadelha and D.~L. Nedel, \emph{{General unitary TFD
  formulation for superstrings}},
  \href{https://doi.org/10.22323/1.013.0032}{\emph{PoS} {\bfseries WC2004}
  (2004) 032} [\href{https://arxiv.org/abs/hep-th/0412128}{{\ttfamily
  hep-th/0412128}}].

\bibitem{watson}
E.~T. Whittaker and G.~N. Watson, \emph{A Course of Modern Analysis}. Cambridge
  University Press, 1920.

\bibitem{KimLee}
S.~P. Kim and C.~H. Lee, \emph{{Nonequilibrium quantum dynamics of second order
  phase transitions}},
  \href{https://doi.org/10.1103/PhysRevD.62.125020}{\emph{Phys. Rev. D}
  {\bfseries 62} (2000) 125020}
  [\href{https://arxiv.org/abs/hep-ph/0005224}{{\ttfamily hep-ph/0005224}}].

\bibitem{ume1}
H.~Umezawa, \emph{{Advanced field theory: Micro, macro, and thermal physics}}.
  American Institute of Physics, 1995.

\bibitem{ume2}
Y.~Takahasi and H.~Umezawa, \emph{{Thermo field dynamics}}, {\emph{Collect.
  Phenom.} {\bfseries 2} (1975) 55}.

\bibitem{adsnos}
M.~B. Cantcheff, A.~L. Gadelha, D.~F.~Z. Marchioro and D.~L. Nedel,
  \emph{{String in AdS Black Hole: A Thermo Field Dynamic Approach}},
  \href{https://doi.org/10.1103/PhysRevD.86.086006}{\emph{Phys. Rev.}
  {\bfseries D86} (2012) 086006}
  [\href{https://arxiv.org/abs/1205.3438}{{\ttfamily 1205.3438}}].

\bibitem{nedel1}
D.~Nedel, M.~Abdalla and A.~Gadelha, \emph{{Superstring in a pp-wave background
  at finite temperature: TFD approach}},
  \href{https://doi.org/10.1016/j.physletb.2004.08.013}{\emph{Phys. Lett. B}
  {\bfseries 598} (2004) 121}
  [\href{https://arxiv.org/abs/hep-th/0405258}{{\ttfamily hep-th/0405258}}].

\bibitem{torus}
M.~Abdalla, A.~Gadelha and D.~L. Nedel, \emph{{Closed string thermal torus from
  thermofield dynamics}},
  \href{https://doi.org/10.1016/j.j.physletb.2005.03.048}{\emph{Phys. Lett. B}
  {\bfseries 613} (2005) 213}
  [\href{https://arxiv.org/abs/hep-th/0410068}{{\ttfamily hep-th/0410068}}].

\bibitem{nedel2}
M.~Abdalla, A.~Gadelha and D.~L. Nedel, \emph{{PP-wave light-cone free string
  field theory at finite temperature}},
  \href{https://doi.org/10.1088/1126-6708/2005/10/063}{\emph{JHEP} {\bfseries
  10} (2005) 063} [\href{https://arxiv.org/abs/hep-th/0508195}{{\ttfamily
  hep-th/0508195}}].

\bibitem{BEY}
R.~Brustein, M.~B. Einhorn and A.~Yarom, \emph{{Entanglement and Nonunitary
  Evolution}}, \href{https://doi.org/10.1088/1126-6708/2007/04/086}{\emph{JHEP}
  {\bfseries 04} (2007) 086}
  [\href{https://arxiv.org/abs/hep-th/0609075}{{\ttfamily hep-th/0609075}}].

\bibitem{garvit}
P.~Garbaczewski and G.~Vitiello, \emph{{A Canonical Description of the Solitary
  Quantum Decay}}, {\emph{Nuovo Cim.} {\bfseries A44} (1978) 108}.

\bibitem{ceravi}
E.~Celeghini, M.~Rasetti and G.~Vitiello, \emph{{Quantum dissipation}},
  \href{https://doi.org/10.1016/0003-4916(92)90302-3}{\emph{Annals Phys.}
  {\bfseries 215} (1992) 156}.

\bibitem{ChuUme}
H.~Chu and H.~Umezawa, \emph{{A Unified formalism of thermal quantum field
  theory}}, \href{https://doi.org/10.1142/S0217751X94000960}{\emph{Int. J. Mod.
  Phys.} {\bfseries A9} (1994) 2363}.

\bibitem{lands}
N.~P. Landsman and C.~G. van Weert, \emph{{Real and Imaginary Time Field Theory
  at Finite Temperature and Density}},
  \href{https://doi.org/10.1016/0370-1573(87)90121-9}{\emph{Phys. Rept.}
  {\bfseries 145} (1987) 141}.

\bibitem{Madhu}
K.~Madhu and K.~Narayan, \emph{{String spectra near some null cosmological
  singularities}},
  \href{https://doi.org/10.1103/PhysRevD.79.126009}{\emph{Phys. Rev. D}
  {\bfseries 79} (2009) 126009}
  [\href{https://arxiv.org/abs/0904.4532}{{\ttfamily 0904.4532}}].

\bibitem{dafdani08}
D.~Z. Marchioro and D.~L. Nedel, \emph{{Observer dependent D-brane for strings
  propagating in pp-wave time dependent background}},
  \href{https://doi.org/10.1140/epjc/s10052-008-0569-7}{\emph{Eur. Phys. J. C}
  {\bfseries 55} (2008) 343} [\href{https://arxiv.org/abs/0711.0556}{{\ttfamily
  0711.0556}}].

\bibitem{Lash}
N.~Lashkari, A.~Dymarsky and H.~Liu, \emph{{Eigenstate Thermalization
  Hypothesis in Conformal Field Theory}},
  \href{https://doi.org/10.1088/1742-5468/aab020}{\emph{J. Stat. Mech.}
  {\bfseries 1803} (2018) 033101}
  [\href{https://arxiv.org/abs/1610.00302}{{\ttfamily 1610.00302}}].

\end{thebibliography}\endgroup

\end{document}